\documentclass[a4paper,11pt]{article}

\usepackage{amsmath,amssymb,graphicx,enumerate,lipsum}
\usepackage[margin=2cm]{geometry}
\usepackage{framed,color}
\definecolor{LightGray}{rgb}{0.93,0.93,0.93}
\definecolor{Gray}{rgb}{0.4,0.4,0.4}
\definecolor{DarkBlue}{rgb}{0,0,0.2}
\usepackage[colorlinks = true,allcolors=blue]{hyperref}
\usepackage[size=small]{caption}
\usepackage[numbers,sort&compress]{natbib}
\usepackage{bold-extra}

\newcommand{\algbox}[1]{\vspace{0.0cm}\fcolorbox{LightGray}{LightGray}{\parbox{0.99\textwidth}{\color{black}\vspace{-0.2cm}#1 \vspace{-0.2cm}}}\vspace{0.2cm}}
\newtheorem{algorithm}{Algorithm}[section]
\newcommand{\comment}[1]{\textcolor{Gray}{\% #1}}
\newcommand{\subhead}[1]{\noindent\textcolor{black}{\bfseries\textsc{#1}}}

\renewcommand{\thealgorithm}{\arabic{algorithm}}
\title{Stochastic models of advection-diffusion in layered media}

\author{Elliot J. Carr\footnote{Corresponding Author (\href{elliot.carr@qut.edu.au}{elliot.carr@qut.edu.au})}\\ \small School of Mathematical Sciences, Queensland University of Technology (QUT), Brisbane, Australia}

\date{}

\begin{document}
\maketitle

\begin{abstract}
Mathematically modelling diffusive and advective transport of particles in heterogeneous layered media is important to many applications in computational, biological and medical physics. While deterministic continuum models of such transport processes are well established, they fail to account for randomness inherent in many problems and are valid only for a large number of particles. To address this, this paper derives a suite of equivalent stochastic (discrete-time discrete-space random walk) models for several standard continuum (partial differential equation) models of diffusion and advection-diffusion across a fully- or semi-permeable interface. Our approach involves discretising the continuum model in space and time to yield a Markov chain, which governs the transition probabilities between spatial lattice sites during each time step. Discretisation in space is carried out using a standard finite volume method while two options are considered for discretisation in time. A simple forward Euler discretisation yields a stochastic model taking the form of a local (nearest-neighbour) random walk with simple analytical expressions for the transition probabilities while an exact exponential discretisation yields a non-local random walk with transition probabilities defined numerically via a matrix exponential. Constraints on the size of the spatial and/or temporal steps are provided for each option to ensure the transition probabilities are non-negative. MATLAB code comparing the stochastic and continuum models is available on GitHub (\href{https://github.com/elliotcarr/Carr2024c}{https://github.com/elliotcarr/Carr2024c}) with simulation results demonstrating good agreement for several example problems.
\end{abstract}

\section{Introduction}
The transport of ``particles'' (e.g. cells, molecules, etc) via diffusion (unbiased or undirected movement) and advection (biased or directed movement) is fundamental to computational, biological and medical physics. A key challenge when modelling such transport processes is the presence of heterogeneity in the medium, where the rates of diffusion and/or advection vary spatially. Arising frequently in applications and the focus of this paper is the layered medium, where homogeneous layers are separated by interfaces or barriers. Examples include drug delivery via multilayered capsules \cite{carr_2018,kaoui_2018,hadjitheodorou_2014}, cell movement or heat transport in layered biological tissue \cite{mantzavinos_2016,mcinerney_2019,becker_2013} and contaminant transport in aquifer layers \cite{carr_2020,liu_1998,chen_2016}. In such applications, mathematical models provide the ability to make predictions, improve the understanding of the governing processes and investigate the effect of parameters. 

Mathematical models of particle transport typically take one of two forms: deterministic continuum models or stochastic discrete models. Continuum models consider the particles as a collective (continuum) described by the particle density, a continuous function of space and time, with transport governed by a partial differential equation. Stochastic models instead consider the particles individually described by their individual positions, discrete points in space and time, with the transport of each individual particle governed by probabilistic rules. Continuum models are attractive because they can employ the analysis of continuous mathematics to extract insight, however, their validity relies on a large number of particles \cite{gavagnin_2018}. Discrete stochastic models are attractive because they account for the randomness inherent in many physical processes and remain valid for a small number of particles, however, they can be computationally expensive for a large number of particles \cite{gavagnin_2018}. Due to these advantages and disadvantages, it is important to determine equivalent continuum and stochastic models to allow the appropriate model to be used depending on the application.

Generally speaking, there are two approaches for developing consistent continuum and stochastic models of particle transport. The first approach begins by writing down the stochastic model (phenomenological probabilistic rules governing individual particle movement) and then use Taylor series expansions to derive a continuum (partial differential equation) model for the particle density \cite{codling_2008,ibe_2013,redner_2001}. Alternatively, one can start by writing down the continuum model for the particle density (a partial differential equation with initial and boundary conditions) and discretise in space and time to derive the stochastic model (list of transition probabilities between lattice sites) for the movement of individual particles \cite{cai_2006,anderson_1998,meinecke_2016}. For homogeneous media (spatially constant rates of diffusion and advection), the relationship between the transition probabilities in the stochastic model and parameters in the continuum model are well known \cite{codling_2008,ibe_2013,carr_2019,ellery_2012,redner_2001}. Heterogeneous media (spatially variable rates of diffusion and advection), however, presents a key challenge. In the first approach, determining a continuum model given a set of probabilistic rules governing individual particle movement can be difficult \cite{gavagnin_2018}, while in the second approach, discretising the continuum model can lead to negative transition probabilities if care is not taken \cite{meinecke_2016,lotstedt_2015,isaacson_2018}.

Continuum models of diffusion or advection-diffusion in layered media are well established~\cite{carr_2020,alemany_2022,sheils_2017,carr_2016,guerrero_2013,liu_1998,hickson_2009}. Here, homogeneous diffusion or advection-diffusion equations on each layer are coupled by internal boundary conditions between adjacent layers that describe either fully permeable (no resistance to particle transport between adjacent layers) or semi-permeable (resistance to particle transport between adjacent layers) interfaces. In this paper, we derive a suite of equivalent stochastic (discrete-time discrete-space random walk) models for several well-established continuum models of diffusion and advection-diffusion across a fully- or semi-permeable interface. Linear continuum models are considered which give rise to stochastic models involving non-interacting particles that move independently of one another~\cite{simpson_2009}. Our approach involves discretising the continuum model in space and time to yield a Markov chain that governs the transition of particles in space and time. Discretisation in space is carried out using a standard finite volume method while two options are considered for discretisation in time. A forward Euler discretisation in time yields a stochastic model taking the form of a local (nearest-neighbour) random walk with simple analytical expressions for the transition probabilities while an exact exponential discretisation in time yields a non-local random walk with transition probabilities defined numerically via a matrix exponential. Both stochastic models are distinct from those reported in recent work \cite{alemany_2022,das_2023} and accurately mimic the behaviour of the particle density obtained from the continuum model.

The remaining sections of this paper are organised as follows. In the next section, we outline the different continuum models considered in this work. In section \ref{sec:stochastic_models}, we describe the general approach employed for deriving an equivalent stochastic model and present the main results of the paper: equivalent stochastic models for each of the continuum models together with simulations confirming agreement. The work is then summarised and concluded in section \ref{sec:conclusions}.

\begin{table*}[p]
\setlength{\fboxsep}{0.8em}
\noindent\fbox{\begin{minipage}{0.96\textwidth}

Model 1: Homogeneous diffusion
\begin{gather*}
\frac{\partial u}{\partial t} = D\frac{\partial^{2} u}{\partial x^{2}},\quad 0 < x < L,\quad\\
u(x,0) = f(x),\quad \frac{\partial u}{\partial x}(0,t) = 0,\quad \frac{\partial u}{\partial x}(L,t) = 0.
\end{gather*}

\medskip
Model 2: Heterogeneous diffusion with fully-permeable interface
\begin{gather*}
\frac{\partial u_{1}}{\partial t} = D_{1}\frac{\partial^{2} u_{1}}{\partial x^{2}},\quad 0 < x < \ell,\quad
\frac{\partial u_{2}}{\partial t} = D_{2}\frac{\partial^{2} u_{2}}{\partial x^{2}},\quad \ell < x < L,\\
u_{1}(\ell,t) = u_{2}(\ell,t),\quad D_{1}\frac{\partial u_{1}}{\partial x}(\ell,t) = D_{2}\frac{\partial u_{2}}{\partial x}(\ell,t),\\
u_{1}(x,0) = f(x),\quad u_{2}(x,0) = f(x),\quad \frac{\partial u_{1}}{\partial x}(0,t) = 0,\quad \frac{\partial u_{2}}{\partial x}(L,t) = 0.
\end{gather*}

\medskip
Model 3: Heterogeneous diffusion with semi-permeable interface
\begin{gather*}
\frac{\partial u_{1}}{\partial t} = D_{1}\frac{\partial^{2} u_{1}}{\partial x^{2}},\quad 0 < x < \ell,\quad
\frac{\partial u_{2}}{\partial t} = D_{2}\frac{\partial^{2} u_{2}}{\partial x^{2}},\quad \ell < x < L,\\
D_{1}\frac{\partial u_{1}}{\partial x}(\ell,t) = D_{2}\frac{\partial u_{2}}{\partial x}(\ell,t) = H[u_{2}(\ell,t)-u_{1}(\ell,t)],\\
u_{1}(x,0) = f(x),\quad u_{2}(x,0) = f(x),\quad \frac{\partial u_{1}}{\partial x}(0,t) = 0,\quad \frac{\partial u_{2}}{\partial x}(L,t) = 0.
\end{gather*}

\medskip
Model 4: Homogeneous advection-diffusion
\begin{gather*}
\frac{\partial u}{\partial t} = D\frac{\partial^{2} u}{\partial x^{2}} - v\frac{\partial u}{\partial x},\quad 0 < x < L,\quad\\
u(x,0) = f(x),\quad vu(0,t) - D\frac{\partial u}{\partial x}(0,t) = 0,\quad vu(L,t) - D\frac{\partial u}{\partial x}(L,t) = 0.
\end{gather*}

\medskip
Model 5: Heterogeneous advection-diffusion with fully-permeable interface
\begin{gather*}
\frac{\partial u_{1}}{\partial t} = D_{1}\frac{\partial^{2} u_{1}}{\partial x^{2}} - v_{1}\frac{\partial u_{1}}{\partial x},\quad 0 < x < \ell,\quad
\frac{\partial u_{2}}{\partial t} = D_{2}\frac{\partial^{2} u_{2}}{\partial x^{2}} - v_{2}\frac{\partial u_{2}}{\partial x},\quad \ell < x < L,\\
u_{1}(\ell,t) = u_{2}(\ell,t),\quad v_{1}u_{1}(\ell,t) - D_{1}\frac{\partial u_{1}}{\partial x}(\ell,t) = v_{2}u_{2}(\ell,t) - D_{2}\frac{\partial u_{2}}{\partial x}(\ell,t),\\
u_{1}(x,0) = f(x),\quad u_{2}(x,0) = f(x),\quad v_{1}u_{1}(0,t) - D_{1}\frac{\partial u_{1}}{\partial x}(0,t) = 0,\quad v_{2}u_{2}(L,t) - D_{2}\frac{\partial u_{2}}{\partial x}(L,t) = 0.
\end{gather*}

\medskip
Model 6: Heterogeneous advection-diffusion with semi-permeable interface
\begin{gather*}
\frac{\partial u_{1}}{\partial t} = D_{1}\frac{\partial^{2} u_{1}}{\partial x^{2}} - v_{1}\frac{\partial u_{1}}{\partial x},\quad 0 < x < \ell,\quad
\frac{\partial u_{2}}{\partial t} = D_{2}\frac{\partial^{2} u_{2}}{\partial x^{2}} - v_{2}\frac{\partial u_{2}}{\partial x},\quad \ell < x < L,\\
D_{1}\frac{\partial u_{1}}{\partial x}(\ell,t) - v_{1}u_{1}(\ell,t) = D_{2}\frac{\partial u_{2}}{\partial x}(\ell,t) - v_{2}u_{2}(\ell,t) = H[u_{2}(\ell,t)-u_{1}(\ell,t)],\\
u_{1}(x,0) = f(x),\quad u_{2}(x,0) = f(x),\quad v_{1}u_{1}(0,t) - D_{1}\frac{\partial u_{1}}{\partial x}(0,t) = 0,\quad v_{2}u_{2}(L,t) - D_{2}\frac{\partial u_{2}}{\partial x}(L,t) = 0.
\end{gather*}
\end{minipage}}
\caption{Continuum models considered in this paper (see section \ref{sec:continuum_models} for further details). Stochastic models for each of the above continuum models are derived in section \ref{sec:stochastic_models}.}
\label{tab:continuum_models}
\end{table*} 

\section{Continuum Models}
\label{sec:continuum_models}
\begin{figure}[p]
\centering
\def\figw{0.45\textwidth}
\includegraphics[scale=0.95]{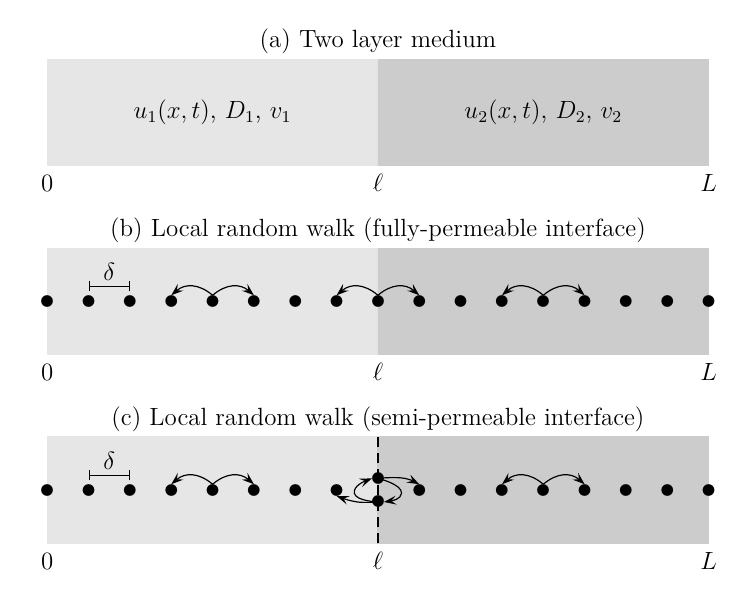}\\[1em]
\includegraphics[width=\figw]{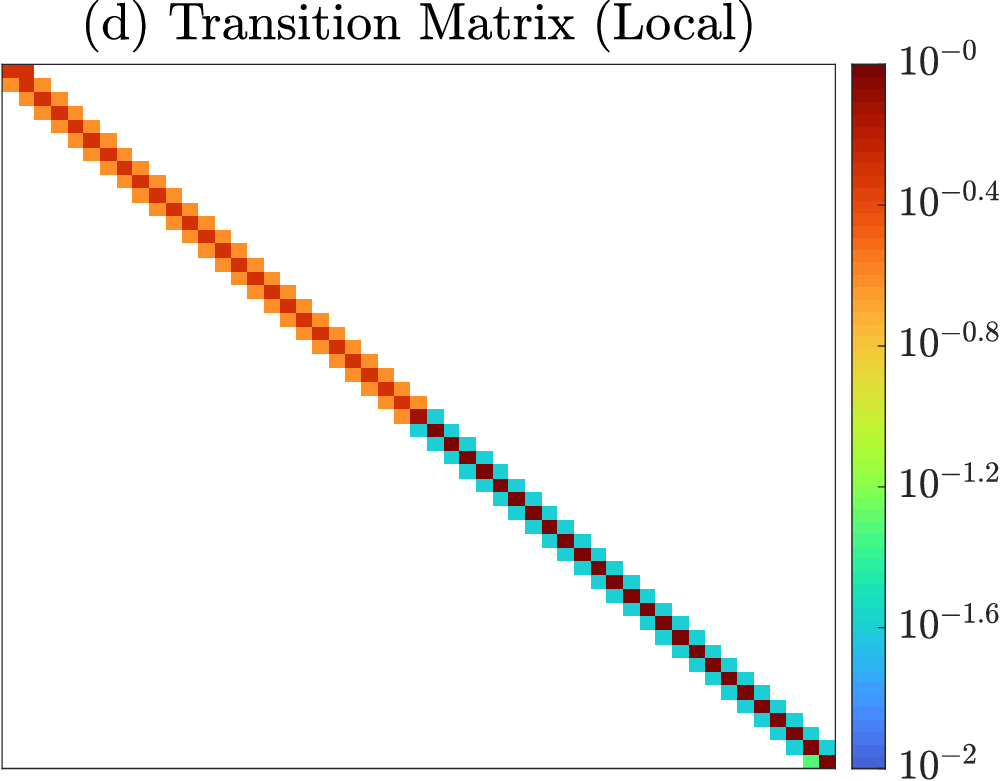}\hspace{0.1\textwidth}\includegraphics[width=\figw]{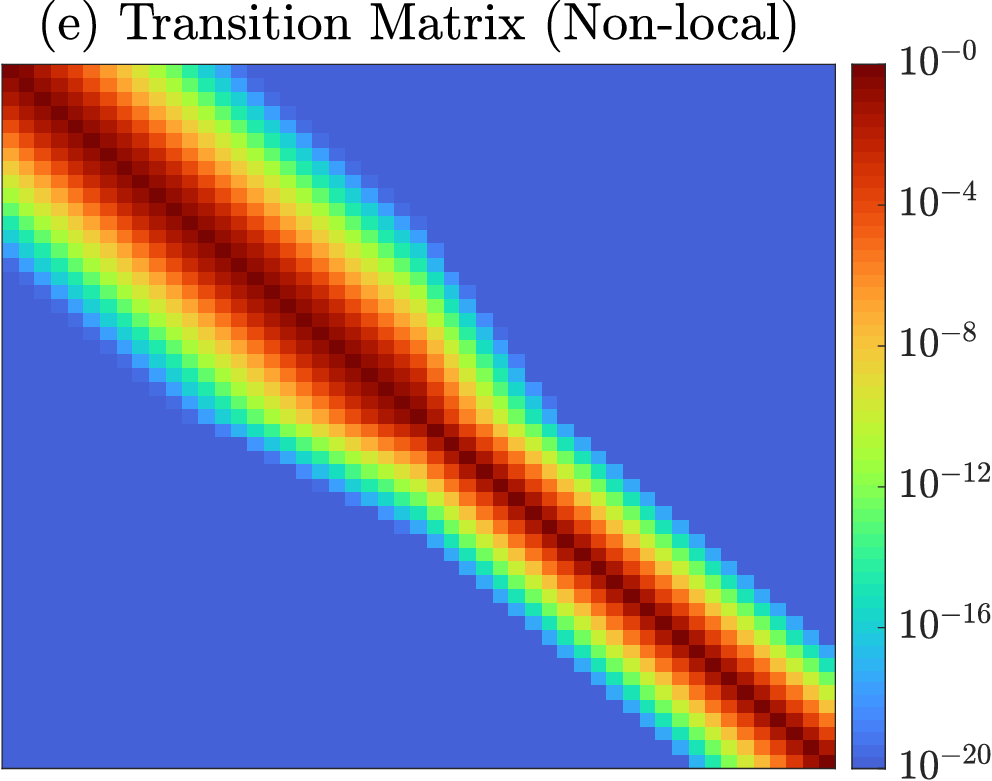}\\[1em]
\caption{(a) Heterogeneous (two layer) medium with piecewise particle density, diffusivity and advective velocity: $u_{1}(x,t)$, $D_{1}$ and $v_{1}$ in the first layer ($x\in(0,\ell)$) and $u_{2}(x,t)$, $D_{2}$ and $v_{2}$ in the second layer ($x\in(\ell,L)$). (b)--(c) Local random walk arising from the forward Euler discretisation, where particles can move to neighbouring sites only (or not move) during a single time step. For the non-local random walk arising from the exact exponential discretisation, particles can (in principle) move to any other site (or not move) during a single time step. (b) considers a fully-permeable interface where a single lattice site is located at the interface while (c) considers a semi-permeable interface where two lattice sites are located at the interface (one associated with each layer) to capture the discontinuity in particle density (the two sites are displaced vertically for visual effect only) (d) example tridiagonal transition matrix for the local random walk arising from the forward Euler discretisation (e) example dense transition matrix for the non-local random walk arising from the exact exponential discretisation. (d) and (e) correspond to model 2 with parameters: $N_{s} = 51$, $D_{1} = 0.1$, $D_{2} = 0.01$, $\delta = 0.02$, $\tau = 0.001$. }
\label{fig:stochastic_models}
\end{figure}

We consider the six continuum models outlined in Table \ref{tab:continuum_models}. Models 1--3 consider diffusion only while models 4--6 consider advection-diffusion. Models 1 and 4 consider a homogeneous medium extending from $x=0$ to $x=L$ while models 2, 3, 5 and 6 consider a heterogeneous medium consisting of two layers, the first layer extending from $x=0$ to an interface at $x=\ell$ and the second layer extending from $x=\ell$ to $x=L$ (Figure \ref{fig:stochastic_models}(a)). The particle density in models 1 and 4 is denoted by $u(x,t)$ for all $x\in[0,L]$ while the particle density in models 2, 3, 5 and 6 is denoted by $u_{1}(x,t)$ in the first layer ($x\in[0,\ell]$) and and $u_{2}(x,t)$ in the second layer ($x\in[\ell,L]$). Models 2 and 5 consider a fully-permeable interface with a continuous particle density across the interface while models 3 and 6 consider a semi-permeable interface with a discontinuous particle density across the interface controlled by the transfer coefficient $H$. For models 1 and 4, $D$ defines the diffusivity while for models 2, 3, 5 and 6, $D_{1}$ and $D_{2}$ define the diffusivity in the first and second layers, respectively. For model 4, $v$ defines the advection velocity while for models 5 and 6, $v_{1}$ and $v_{2}$ define the advection velocity in the first and second layers, respectively. In all models, the time interval is taken as finite ($t\in[0,T]$), zero flux (reflecting) boundary conditions are applied at the two external boundaries of the medium ($x=0$ and $x = L$), and the initial particle density is described by the function $f(x)$.

\section{Stochastic Models}
\label{sec:stochastic_models}
For each of the continuum models outlined in Table \ref{tab:continuum_models}, we derive an equivalent stochastic model by discretising the governing equations in space and time. In all cases, we use constant temporal and spatial steps, respectively: $\tau = T/N_{t} >0$ and $\delta = L/(N_{x}-1) > 0$, where $N_{x}$ and $N_{t}$ are positive integers. For models 2, 3, 5 and 6, $N_{x}$ must be chosen so that $\delta$ divides evenly into $\ell$, that is, $\ell/\delta$ is an integer. This choice means that lattice site $\mathcal{I} = \ell/\delta + 1$ is located at the interface ($x=\ell)$. For models 2 and 5, only one lattice site is positioned at the interface as the particle density is continuous ($u_{1}(\ell,t) = u_{2}(\ell,t)$, see Table \ref{tab:continuum_models}), giving $N_{s}=N_{x}$ total sites with site $i$ located at $x = x_{i} := (i-1)\delta$ for $i = 1,\hdots,N_{s}$. For models 3 and 6, two lattice sites are positioned at the interface to capture the discontinuity in the density ($u_{1}(\ell,t) \neq u_{2}(\ell,t)$, see Table \ref{tab:continuum_models}), giving $N_{s}=N_{x}+1$ total sites with site $i$ located at $x = x_{i}$ for $i = 1,\hdots,N_{s}$, where $x_{i} = (i-1)\delta$ for $i = 1,\hdots,\mathcal{I}$ and $x_{i} = (i-2)\delta$ for $i = \mathcal{I}+1,\hdots,N_{s}$. Discretisation in space is carried out using a standard vertex-centered finite volume method \cite{march_2019,carr_2020} with the finite volume for site $i$ defined by the interval $[w_{i},e_{i}]$ where $w_{i} = \max(0,(x_{i-1}+x_{i})/2)$ and $e_{i} = \min((x_{i}+x_{i+1})/2,L)$. Central differences are used to discretise the diffusion term and averaging is used to discretise the advection term \cite{march_2019,carr_2020}. 

For each continuum model in Table \ref{tab:continuum_models}, the discretisation in space yields a system of differential equations (see Appendix \ref{app:fvm_discretisations}):
\begin{gather}
\label{eq:ode_U}
\frac{\text{d}\mathbf{U}}{\text{d}t} = \mathbf{A}\mathbf{U},\qquad \mathbf{U}(0) = \mathbf{U}_{0},
\end{gather}
where $\mathbf{U} = (U_{1},\hdots,U_{N_{s}})^{\mathsf{T}}$, $\mathbf{U}_{0} = (f(x_{1}),\hdots,f(x_{N_{s}}))^{\mathsf{T}}$ and $\mathbf{A}$ is an $N_{s}$ by $N_{s}$ tridiagonal matrix with $U_{i}$ denoting the particle density at lattice site $i$. Equation (\ref{eq:ode_U}) describes the evolution of the particle density at each lattice site over time. To determine the transition probabilities between lattice sites, we first need to convert the system into one involving the number of particles in each control volume. This is achieved using the relationship $U_{i} = N_{i}/(S_{p}V_{i})$, where $N_{i}$ is the number of particles within control volume $i$ (assumed to be at site $i$), $V_{i} = e_{i}-w_{i}$ is the length of control volume $i$ and $S_{p} = \sum_{i=1}^{N_{s}}f(x_{i})V_{i}/N_{p}$ is a scaling constant. Substituting this relationship into (\ref{eq:ode_U}) gives:
\begin{gather}
\label{eq:ode_N}
\frac{\text{d}\mathbf{N}}{\text{d}t} = \mathbf{B}\mathbf{N},\qquad \mathbf{N}(0) = \mathbf{N}_{0},
\end{gather}
where $\mathbf{N} = (N_{1},\hdots,N_{N_{s}})^{\mathsf{T}}$, $\mathbf{N}_{0} = \text{round}(S_{p}(f(x_{1})V_{1},\hdots,f(x_{N_{s}})V_{N_{s}})^{\mathsf{T}})$ and $\mathbf{B} = \mathbf{V}\mathbf{A}\mathbf{V}^{-1}$ is an $N_{s}$ by $N_{s}$ tridiagonal matrix with $\mathbf{V} = \text{diag}(V_{1},\hdots,V_{N_{s}})$ and $\mathbf{V}^{-1} = \text{diag}(V_{1}^{-1},\hdots,V_{N_{s}}^{-1})$. Note that the initial number of particles at each lattice site is rounded to ensure integer numbers.

Discretising (\ref{eq:ode_N}) in time using a one-step method with fixed time step size $\tau$, yields:
\begin{gather}
\label{eq:markov_chain}
\mathbf{N}_{n}^{\mathsf{T}} = \mathbf{N}_{n-1}^{\mathsf{T}}\mathbf{P},\qquad n = 1,\hdots,N_{t},
\end{gather}
where $\mathbf{N}_{k}^{\mathsf{T}}$ denotes $\mathbf{N}^{\mathsf{T}} = (N_{1},\hdots,N_{N_{s}})$ after $k$ time steps (at $t = k\tau$) and $\mathbf{P}$ is an $N_{s}$ by $N_{s}$ mapping matrix that depends on the time discretisation method employed. A key observation is that (\ref{eq:markov_chain}) can be interpreted as a Markov chain provided $\mathbf{P}$ is a (right) stochastic matrix (each row of $\mathbf{P}$ sums to one and each entry of $\mathbf{P}$ is non-negative). In this case, $\mathbf{P}$ defines a matrix of transition probabilities with $p_{i,j}$ (entry of $\mathbf{P}$ in row $i$ and column $j$) providing the probability that a particle located at site $i$ at time $t = t_{n-1}$ moves to site $j$ at time $t = t_{n}$ (note when $i = j$ the particle does not move and remains at site $i$). 

In what follows, we let $\mathbf{C} = \mathbf{B}^{\mathsf{T}} = (\mathbf{V}\mathbf{A}\mathbf{V}^{-1})^{\mathsf{T}}$ and note that for all six continuum models in Table \ref{sec:continuum_models}, $\mathbf{C}$ is tridiagonal and each row of $\mathbf{C}$ sums to zero (see sections \ref{sec:model1}--\ref{sec:model6}). In this work, we consider two distinct time-discretisation methods applied to the system of differential equations (\ref{eq:ode_N}) each yielding a different matrix $\mathbf{P}$: 
\begin{enumerate}[(i)] 
\item \emph{Forward Euler discretisation}: $\mathbf{P} = \mathbf{I}+\tau\mathbf{C}$ where $\mathbf{I}$ is the $N_{s}$ by $N_{s}$ identity matrix, which arises when discretising (\ref{eq:ode_N}) using the forward Euler method from $t=t_{n-1}$ and $t=t_{n}$,
\item \emph{Exact exponential discretisation}: $\mathbf{P} = \exp(\tau\mathbf{C})$, where $\exp(\cdot)$ is the matrix exponential, which arises when solving (\ref{eq:ode_N}) exactly from $t=t_{n-1}$ and $t=t_{n}$.
\end{enumerate}
The exponential Euler discretisation yields transition probabilities defined numerically via the matrix exponential while the forward Euler discretisation yields simple analytical expressions for the transition probabilities (as we will see later in sections \ref{sec:model1}--\ref{sec:model6}).

Both the forward Euler discretisation and exact exponential discretisation yield a right stochastic matrix $\mathbf{P}$ and hence a Markov chain (\ref{eq:markov_chain}). For the forward Euler discretisation, each row of $\mathbf{P} = \mathbf{I}+\tau\mathbf{C}$ sums to one since each row of $\mathbf{C}$ sums to zero but constraints on the time step $\tau$ (models 1--6) and spatial step $\delta$ (models 4--6) are required to ensure each entry of $\mathbf{P}$ is non-negative (see sections \ref{sec:model1}--\ref{sec:model6}). For the exact exponential discretisation, each row of $\mathbf{P} = \exp(\tau\mathbf{C}) = \mathbf{I}+\tau\mathbf{C} + \frac{{\tau}^{2}}{2}\mathbf{C}^{2} + \cdots $ sums to one since each row of $\mathbf{C}^{k}$ ($k = 2,3,\hdots$) sums to zero when each row of $\mathbf{C}$ sums to zero but each entry of $\mathbf{P} = \exp(\tau\mathbf{C})$ is non-negative if an only if all off-diagonal entries of $\mathbf{C}$ are positive \cite{higham_2008}, which requires constraints on $\delta$ for the advection-diffusion models (models 4--6) (see sections \ref{sec:model4}--\ref{sec:model6}). 

The constraints required to ensure $\mathbf{P}$ is non-negative are equivalent to the constraints required to ensure monotonicity \cite{holmes_2007} of the corresponding continuum model solution. For the forward Euler discretisation, $\mathbf{P} = \mathbf{I}+\tau\mathbf{C} = \mathbf{V}^{-\!\mathsf{T}}(\mathbf{I}+\tau\mathbf{A})^{\mathsf{T}}\mathbf{V}^{\mathsf{T}}$ has non-negative entries if the entries of $\mathbf{I}+\tau\mathbf{A}$ are non-negative, which is the monotonicity condition for the forward Euler discretisation of (\ref{eq:ode_U}): $\mathbf{U}_{n} = (\mathbf{I}+\tau\mathbf{A})\mathbf{U}_{n-1}$, where $\mathbf{U}_{k}$ denotes $\mathbf{U}$ after $k$ time steps (at $t = k\tau$). Similarly, for the exact exponential discretisation, $\mathbf{P} = \exp(\tau\mathbf{C}) = \mathbf{V}^{-\!\mathsf{T}}\exp(\tau\mathbf{A})^{\mathsf{T}}\mathbf{V}^{\mathsf{T}}$ has non-negative entries provided the entries of $\exp(\tau\mathbf{A})$ are non-negative, which is the monotonicity condition for the exact exponential discretisation of (\ref{eq:ode_U}): $\mathbf{U}_{n} = \exp(\tau\mathbf{A})\mathbf{U}_{n-1}$.

When the above conditions are met, the forward Euler discretisation yields a local random walk model (Figure \ref{fig:stochastic_models}(b)(c)), where particles can move to neighbouring sites only during a single time step (since $\mathbf{P} = \mathbf{I}+\tau\mathbf{C}$ is tridiagonal) while the exact exponential discretisation yields a non-local random walk model, where particles can move to any other site during a single time step (since $\mathbf{P} = \exp(\tau\mathbf{C})$ is dense, albeit with probabilities that become smaller the further the move). Note that the non-local random walk model is consistent with the infinite propagation speed of continuum diffusion: the particle density at every point in $[0,L]$ affects the particle density at every other point in $[0,L]$ \cite{strauss_2008}. Example transition matrices for both the forward Euler and exact exponential discretisations are given in Figure \ref{fig:stochastic_models}(d)(e) for model 2. Here, particles in the first layer are more likely to move than particles in the second layer and particles at the interface are more likely to move to the left than to the right since $D_{1}>D_{2}$ (see section \ref{sec:model2}). The transition matrix for the exact exponential discretisation decays rapidly away from the diagonal with the transition probability effectively zero outside of a band around the diagonal; behaviour that is typical of the exponential of tridiagonal matrices~\cite{iserles_1999,benzi_2015}. 

Given the transition matrix $\mathbf{P}$, the stochastic model governing the movement of particles is described in Algorithm \ref{alg:stochastic_model1}. At each time step, a uniform random number on $(0,1)$ is generated for each particle, with the particle moving from its current lattice site $i$ to a new lattice site $j$ if $r\in (P_{i,j-1},P_{i,j})$ where $P_{i,0} = 0$ and $P_{i,k} = \sum_{m=1}^{k} p_{i,m}$. When the position of each individual particle is required (rather than just the number of particles at each lattice site) an alternative version of the algorithm can be implemented (see Appendix \ref{app:stochastic_model2}) .

\bigskip\noindent
\algbox{
\begin{algorithm}[Stochastic Model]\mbox{}\\
\label{alg:stochastic_model1}
\emph{
\begin{tabular}{@{}l}
$U_{i,0} = f(x_{i})$  for all $i = 1,\hdots,N_{s}$ \comment{particle density at lattice site $i$ and time $t=0$}\\
$S_{p} = \sum_{i=1}^{N_{s}}U_{i,0}V_{i}/N_{p}$ \comment{scaling constant}\\
$N_{i,0} = \text{round}(U_{i,0}V_{i}/S_{p})$ for all $i = 1,\hdots,N_{s}$ \comment{initial number of particles at site $i$}\\
$P_{i,0} = 0$ and $P_{i,j} = \sum_{m=1}^{j} p_{i,m}$ for all $i=1,\hdots,N_{s}$ and $j=1,\hdots,N_{s}$ \comment{cumulative probabilities}\\
\textbf{for} $n = 1,\hdots,N_{t}$ \comment{loop over time steps}\\
\qquad $N_{i,n} = N_{i,n-1}$ for all $i = 1,\hdots,N_{s}$ \comment{number of particles at lattice site $i$ and time $t = t_{n-1}$}\\
\qquad \textbf{for} $i = 1,\hdots,N_{s}$ \comment{loop over lattice sites}\\
\qquad \qquad \textbf{for} $k = 1,\hdots,N_{i,n-1}$ \comment{loop over number of particles at lattice site $i$}\\
\qquad \qquad \qquad Sample $r \sim \text{Uniform}(0,1)$ \comment{uniform random number in $[0,1]$}\\
\qquad \qquad \qquad Find $j$ such that $r\in(P_{i,j-1},P_{i,j})$ \comment{particle moves from site $i$ to site $j$ at $t = t_{n}$}\\
\qquad \qquad \qquad $N_{i,n} = N_{i,n} - 1$, $N_{j,n} = N_{j,n} + 1$ \comment{update number of particles at sites $i$ and $j$}\\ 
\qquad \qquad \textbf{end}\\
\qquad \textbf{end}\\
\qquad $U_{i,n} = N_{i,n}S_{p} / V_{i}$ for all $i = 1,\hdots,N_{s}$ \comment{particle density at lattice site $i$ and time $t = t_{n}$}\\
\textbf{end}\\
\end{tabular}
}
\end{algorithm}}

\bigskip
\noindent In the following subsections, we (i) give the discretisation matrix $\mathbf{C}$ for models 1--6 that defines the transition matrix in the stochastic model (\ref{eq:markov_chain}) (ii) provide analytical expressions for the transition probabilities for the case of the forward Euler discretisation (iii) discuss the transition matrix for the exact exponential discretisation (v) provide any constraints on the time and spatial steps $\tau$ and $\delta$ (both are positive by definition) and (iv) confirm the equivalence of the stochastic and continuum models using example simulations. In what follows below, $c_{i,j}$ denotes the entry of $\mathbf{C}$ in row $i$ and column $j$ while $p_{i,j}$ is the probability a particle moves from lattice site $i$ at time $t = t_{n}$ to lattice site $j$ at time $t = t_{n+1} = t_{n} + \tau$.

\subsection{Model 1: Homogeneous diffusion}
\label{sec:model1}
\subhead{Discretisation Matrix}: The matrix $\mathbf{C}$ has entries:
\begin{gather*}
c_{i,i} = -\frac{2D}{\delta^{2}}, \quad c_{i,i+1} = \frac{2D}{\delta^{2}},\quad i = 1,\\
c_{i,i-1} = \frac{D}{\delta^{2}}, \quad c_{i,i} = -\frac{2D}{\delta^{2}},\quad c_{i,i+1} = \frac{D}{\delta^{2}},\quad i = 2,\hdots,N_{s}-1,\\
c_{i,i-1} = \frac{2D}{\delta^{2}}, \quad c_{i,i} = -\frac{2D}{\delta^{2}},\quad i = N_{s}.
\end{gather*}

\subhead{Forward Euler Discretisation}: The transition matrix $\mathbf{P} = \mathbf{I} + \tau\mathbf{C}$ has entries:
\begin{gather*}
p_{i,i} = 1-\frac{2D\tau}{\delta^{2}}, \quad p_{i,i+1} = \frac{2D\tau}{\delta^{2}},\quad i = 1,\\
p_{i,i-1} = \frac{D\tau}{\delta^{2}}, \quad p_{i,i} = 1-\frac{2D\tau}{\delta^{2}},\quad p_{i,i+1} = \frac{D\tau}{\delta^{2}},\quad i = 2,\hdots,N_{s}-1,\\
p_{i,i-1} = \frac{2D\tau}{\delta^{2}}, \quad p_{i,i} = 1-\frac{2D\tau}{\delta^{2}},\quad i = N_{s},
\end{gather*}
and is a stochastic matrix when $\tau \leq \frac{\delta^{2}}{2D}$ with no restriction on $\delta$. Inspecting the transition probabilities, we see that particles move with probability $P_{m} = \frac{2D\tau}{\delta^{2}}$ or remain at their current site with probability $1 - P_{m}$, which agrees with the classical formula, $D = \frac{P_{m}\delta^{2}}{2\tau}$, obtained from Taylor series expansion of the stochastic model \cite{redner_2001,codling_2008}. At the interior sites ($i=2,\hdots,N_{s}-1$), particles move to the left and right with equal probability, $P_{l} = P_{r} = \frac{D\tau}{\delta^{2}}$, with movement occurring more frequently for larger values of $D$. \bigskip

\subhead{Exact Exponential Discretisation}: Each row of $\mathbf{C}$ sums to zero and all off-diagonal entries of $\mathbf{C}$ are non-negative. Therefore, the transition matrix $\mathbf{P} = \exp(\tau\mathbf{C})$ is always a stochastic matrix with no restrictions on $\tau$ and $\delta$.\bigskip 

\subsection{Model 2: Heterogeneous diffusion with fully-permeable interface}
\label{sec:model2}
\subhead{Discretisation Matrix}: The matrix $\mathbf{C}$ has entries:
\begin{gather*}
c_{i,i} = -\frac{2D_{1}}{\delta^{2}}, \quad c_{i,i+1} = \frac{2D_{1}}{\delta^{2}},\quad i = 1,\\
c_{i,i-1} = \frac{D_{1}}{\delta^{2}}, \quad c_{i,i} = -\frac{2D_{1}}{\delta^{2}},\quad c_{i,i+1} = \frac{D_{1}}{\delta^{2}},\quad i = 2,\hdots,\mathcal{I}-1,\\
c_{i,i-1} = \frac{D_{1}}{\delta^{2}},\quad c_{i,i} = -\frac{(D_{1}+D_{2})}{\delta^{2}},\quad c_{i,i+1} = \frac{D_{2}}{\delta^{2}},\quad i = \mathcal{I},\\
c_{i,i-1} = \frac{D_{2}}{\delta^{2}}, \quad c_{i,i} = -\frac{2D_{2}}{\delta^{2}},\quad c_{i,i+1} = \frac{D_{2}}{\delta^{2}},\quad i = \mathcal{I}+1,\hdots,N_{s}-1,\\
c_{i,i-1} = \frac{2D_{2}}{\delta^{2}}, \quad c_{i,i} = -\frac{2D_{2}}{\delta^{2}},\quad i = N_{s}.
\end{gather*}

\subhead{Forward Euler Discretisation}: The transition matrix $\mathbf{P} = \mathbf{I} + \tau\mathbf{C}$ has entries:
\begin{gather*}
p_{i,i} = 1-\frac{2D_{1}\tau}{\delta^{2}}, \quad p_{i,i+1} = \frac{2D_{1}\tau}{\delta^{2}},\quad i = 1,\\
p_{i,i-1} = \frac{D_{1}\tau}{\delta^{2}}, \quad p_{i,i} = 1-\frac{2D_{1}\tau}{\delta^{2}},\quad p_{i,i+1} = \frac{D_{1}\tau}{\delta^{2}},\quad i = 2,\hdots,\mathcal{I}-1,\\
p_{i,i-1} = \frac{D_{1}\tau}{\delta^{2}},\quad p_{i,i} = 1-\frac{(D_{1}+D_{2})\tau}{\delta^{2}},\quad p_{i,i+1} = \frac{D_{2}\tau}{\delta^{2}},\quad i = \mathcal{I},\\
p_{i,i-1} = \frac{D_{2}\tau}{\delta^{2}}, \quad p_{i,i} = 1-\frac{2D_{2}\tau}{\delta^{2}},\quad p_{i,i+1} = \frac{D_{2}\tau}{\delta^{2}},\quad i = \mathcal{I}+1,\hdots,N_{s}-1,\\
p_{i,i-1} = \frac{2D_{2}\tau}{\delta^{2}}, \quad p_{i,i} = 1-\frac{2D_{2}\tau}{\delta^{2}},\quad i = N_{s},
\end{gather*}
and is a stochastic matrix when $\smash{\tau \leq \min\left\{\frac{\delta^{2}}{2D_{1}},\frac{\delta^{2}}{2D_{2}}\right\}}$ with no restriction on $\delta$. Inspecting the transition probabilities, we see that particles move to the left and right with equal probabilities within each layer, $P_{l}=P_{r} = \frac{D_{1}\tau}{\delta^{2}}$ in the first layer and $P_{l}=P_{r}=\frac{D_{2}\tau}{\delta^{2}}$ in the second layer, so particles move more frequently in the first layer than the second layer if $D_{1}>D_{2}$ (or vice versa if $D_{1}<D_{2}$). Furthermore, the change in diffusivity means particles at the interface move to the left and right with different probabilities, $P_{l}=\frac{D_{1}\tau}{\delta^{2}}$ and $P_{r}=\frac{D_{2}\tau}{\delta^{2}}$, with particles more likely to move to the left if $D_{1}>D_{2}$ or to the right if $D_{1}<D_{2}$. These results are consistent with those previously reported in \cite{carr_2019}, where it was shown using Taylor series expansions that they yield the continuum model (model 2 in Table \ref{tab:continuum_models}). \bigskip

\subhead{Exact Exponential Discretisation}: Each row of $\mathbf{C}$ sums to zero and all off-diagonal entries of $\mathbf{C}$ are non-negative. Therefore, the transition matrix $\mathbf{P} = \exp(\tau\mathbf{C})$ is always a stochastic matrix with no restrictions on $\tau$ and $\delta$. \bigskip

\subsection{Model 3: Heterogeneous diffusion with semi-permeable interface}
\label{sec:model3}
\subhead{Discretisation Matrix}: The matrix $\mathbf{C}$ has entries:
\begin{gather*}
c_{i,i} = -\frac{2D_{1}}{\delta^{2}}, \quad c_{i,i+1} = \frac{2D_{1}}{\delta^{2}},\quad i = 1,\\
c_{i,i-1} = \frac{D_{1}}{\delta^{2}}, \quad c_{i,i} = -\frac{2D_{1}}{\delta^{2}},\quad c_{i,i+1} = \frac{D_{1}}{\delta^{2}},\quad i = 2,\hdots,\mathcal{I}-1,\\
c_{i,i-1} = \frac{2D_{1}}{\delta^{2}},\quad c_{i,i} = -\frac{2(D_{1}+H\delta)}{\delta^{2}},\quad c_{i,i+1} = \frac{2H}{\delta},\quad i = \mathcal{I},\\
c_{i,i-1} = \frac{2H}{\delta},\quad c_{i,i} = -\frac{2(D_{2}+H\delta)}{\delta^{2}},\quad c_{i,i+1} = \frac{2D_{2}}{\delta^{2}},\quad i = \mathcal{I}+1,\\
c_{i,i-1} = \frac{D_{2}}{\delta^{2}}, \quad c_{i,i} = -\frac{2D_{2}}{\delta^{2}},\quad c_{i,i+1} = \frac{D_{2}}{\delta^{2}},\quad i = \mathcal{I}+2,\hdots,N_{s}-1,\\
c_{i,i-1} = \frac{2D_{2}}{\delta^{2}}, \quad c_{i,i} = -\frac{2D_{2}}{\delta^{2}},\quad i = N_{s}.
\end{gather*}
\subhead{Forward Euler Discretisation}: The transition matrix $\mathbf{P} = \mathbf{I} + \tau\mathbf{C}$ has entries:
\begin{gather*}
p_{i,i} = 1-\frac{2D_{1}\tau}{\delta^{2}}, \quad p_{i,i+1} = \frac{2D_{1}\tau}{\delta^{2}},\quad i = 1,\\
p_{i,i-1} = \frac{D_{1}\tau}{\delta^{2}}, \quad p_{i,i} = 1-\frac{2D_{1}\tau}{\delta^{2}},\quad p_{i,i+1} = \frac{D_{1}\tau}{\delta^{2}},\quad i = 2,\hdots,\mathcal{I}-1,\\
p_{i,i-1} = \frac{2D_{1}\tau}{\delta^{2}},\quad p_{i,i} = 1-\frac{2(D_{1}+H\delta)\tau}{\delta^{2}},\quad p_{i,i+1} = \frac{2H\tau}{\delta},\quad i = \mathcal{I},\\
p_{i,i-1} = \frac{2H\tau}{\delta},\quad p_{i,i} = 1-\frac{2(D_{2}+H\delta)\tau}{\delta^{2}},\quad p_{i,i+1} = \frac{2D_{2}\tau}{\delta^{2}},\quad i = \mathcal{I}+1,\\
p_{i,i-1} = \frac{D_{2}\tau}{\delta^{2}}, \quad p_{i,i} = 1-\frac{2D_{2}\tau}{\delta^{2}},\quad p_{i,i+1} = \frac{D_{2}\tau}{\delta^{2}},\quad i = \mathcal{I}+2,\hdots,N_{s}-1,\\
p_{i,i-1} = \frac{2D_{2}\tau}{\delta^{2}}, \quad p_{i,i} = 1-\frac{2D_{2}\tau}{\delta^{2}},\quad i = N_{s},
\end{gather*}
and is a stochastic matrix when $\smash{\tau \leq \min\left\{\frac{\delta^{2}}{2\left(D_{1}+H\delta\right)},\frac{\delta^{2}}{2\left(D_{2}+H\delta\right)}\right\}}$ with no restriction on $\delta$. Inspecting the transition probabilities, we see that the only difference between this model and model 2 occurs at the interface (sites $\mathcal{I}$ and $\mathcal{I}+1$). Here, we see the semi-permeable barrier between the two layers means particles transition between layers (from site $\mathcal{I}$ to site $\mathcal{I}+1$ or vice versa) with probability $\frac{2H\tau}{\delta}$. It follows then that particles transition more frequently between layers for larger values of the transfer coefficient~$H$ and cannot transition between layers when the barrier is impermeable, $H = 0$. \bigskip

\subhead{Exact Exponential Discretisation}: Each row of $\mathbf{C}$ sums to zero and all off-diagonal entries of $\mathbf{C}$ are non-negative. Therefore, the transition matrix $\mathbf{P} = \exp(\tau\mathbf{C})$ is always a stochastic matrix with no restrictions on $\tau$ and $\delta$.

\subsection{Model 4: Homogeneous advection diffusion}
\label{sec:model4}
\subhead{Discretisation Matrix}: The matrix $\mathbf{C}$ has entries:
\begin{gather*}
c_{i,i} = -\frac{(2D+v\delta)}{\delta^{2}}, \quad c_{i,i+1} = \frac{(2D+v\delta)}{\delta^{2}},\quad i = 1,\\
c_{i,i-1} = \frac{(2D-v\delta)}{2\delta^{2}}, \quad c_{i,i} = -\frac{2D}{\delta^{2}},\quad c_{i,i+1} = \frac{(2D+v\delta)}{2\delta^{2}},\quad i = 2,\hdots,N_{s}-1,\\
c_{i,i-1} = \frac{(2D-v\delta)}{\delta^{2}}, \quad c_{i,i} = -\frac{(2D-v\delta)}{\delta^{2}},\quad i = N_{s}.
\end{gather*}

\subhead{Forward Euler Discretisation}: The transition matrix $\mathbf{P} = \mathbf{I} + \tau\mathbf{C}$ has entries:
\begin{gather*}
p_{i,i} = 1-\frac{(2D+v\delta)\tau}{\delta^{2}}, \quad p_{i,i+1} = \frac{(2D+v\delta)\tau}{\delta^{2}},\quad i = 1,\\
p_{i,i-1} = \frac{(2D - v\delta)\tau}{2\delta^{2}}, \quad p_{i,i} = 1-\frac{2D\tau}{\delta^{2}},\quad p_{i,i+1} = \frac{(2D+v\delta)\tau}{2\delta^{2}},\quad i = 2,\hdots,N_{s}-1,\\
p_{i,i-1} = \frac{(2D-v\delta)\tau}{\delta^{2}}, \quad p_{i,i} = 1-\frac{(2D-v\delta)\tau}{\delta^{2}},\quad i = N_{s},
\end{gather*}
and is a stochastic matrix when $\smash{\tau \leq \min\left\{\frac{\delta^{2}}{2D+v\delta},\frac{\delta^{2}}{2D-v\delta}\right\}}$ and $\smash{\delta\leq \frac{2D}{|v|}}$. Inspecting the transition probabilities, we see that inclusion of advection produces biased/directed movement to the right if $v$ is positive and to the left if $v$ is negative, which is consistent with the continuum model (model 4 in Table \ref{tab:continuum_models}). At the interior sites ($i=2,\hdots,N_{s}-1$), particles move to the left and right with probabilities $\smash{P_{l} = \frac{(2D-v\delta)\tau}{2\delta^{2}}}$ and $\smash{P_{r} = \frac{(2D+v\delta)\tau}{2\delta^{2}}}$, which agree with the classical formulas, $\smash{D = \frac{(P_{l}+P_{r})\delta^{2}}{2\tau}}$ and $\smash{v = \frac{\delta(P_{r}-P_{l})}{\tau}}$, obtained from Taylor series expansion of the stochastic model \cite{codling_2008,redner_2001}. \bigskip

\subhead{Exact Exponential Discretisation}: Each row of $\mathbf{C}$ sums to zero while all off-diagonal entries of $\mathbf{C}$ are non-negative when $\smash{\delta\leq \frac{2D}{|v|}}$. Therefore, the transition matrix $\mathbf{P} = \exp(\tau\mathbf{C})$ is a stochastic matrix when $\delta\leq \frac{2D}{|v|}$ with no restriction on $\tau$.

\subsection{Model 5: Heterogeneous advection-diffusion with fully-permeable interface}
\label{sec:model5}

\subhead{Discretisation Matrix}: The matrix $\mathbf{C}$ has entries:
\begin{gather*}
c_{i,i} = -\frac{(2D_{1}+v_{1}\delta)}{\delta^{2}}, \quad c_{i,i+1} = \frac{(2D_{1}+v_{1}\delta)}{\delta^{2}},\quad i = 1,\\
c_{i,i-1} = \frac{(2D_{1} - v_{1}\delta)}{2\delta^{2}}, \quad c_{i,i} = -\frac{2D_{1}}{\delta^{2}},\quad c_{i,i+1} = \frac{(2D_{1}+v_{1}\delta)}{2\delta^{2}},\quad i = 2,\hdots,\mathcal{I}-1,\\
c_{i,i-1} = \frac{(2D_{1} - v_{1}\delta)}{2\delta^{2}}, \quad c_{i,i} = -\frac{[2(D_{1}+D_{2}) + (v_{2}-v_{1})\delta]}{2\delta^{2}},\quad c_{i,i+1} = \frac{(2D_{2}+v_{2}\delta)}{2\delta^{2}},\quad i = \mathcal{I},\\
c_{i,i-1} = \frac{(2D_{2} - v_{2}\delta)}{2\delta^{2}}, \quad c_{i,i} = -\frac{2D_{2}}{\delta^{2}},\quad c_{i,i+1} = \frac{(2D_{2}+v_{2}\delta)}{2\delta^{2}},\quad i = \mathcal{I}+1,\hdots,N_{s}-1,\\\
c_{i,i-1} = \frac{(2D_{2}-v_{2}\delta)}{\delta^{2}}, \quad c_{i,i} = -\frac{(2D_{2}-v_{2}\delta)}{\delta^{2}},\quad i = N_{s}.
\end{gather*}

\subhead{Forward Euler Discretisation}: The transition matrix $\mathbf{P} = \mathbf{I} + \tau\mathbf{C}$ has entries:
\begin{gather*}
p_{i,i} = 1-\frac{(2D_{1}+v_{1}\delta)\tau}{\delta^{2}}, \quad p_{i,i+1} = \frac{(2D_{1}+v_{1}\delta)\tau}{\delta^{2}},\quad i = 1,\\
p_{i,i-1} = \frac{(2D_{1} - v_{1}\delta)\tau}{2\delta^{2}}, \quad p_{i,i} = 1-\frac{2D_{1}\tau}{\delta^{2}},\quad p_{i,i+1} = \frac{(2D_{1}+v_{1}\delta)\tau}{2\delta^{2}},\quad i = 2,\hdots,\mathcal{I}-1,\\
p_{i,i-1} = \frac{(2D_{1} - v_{1}\delta)\tau}{2\delta^{2}}, \quad p_{i,i} = 1-\frac{[2(D_{1}+D_{2}) + (v_{2}-v_{1})\delta]\tau}{2\delta^{2}},\quad p_{i,i+1} = \frac{(2D_{2}+v_{2}\delta)\tau}{2\delta^{2}},\quad i = \mathcal{I},\\
p_{i,i-1} = \frac{(2D_{2} - v_{2}\delta)\tau}{2\delta^{2}}, \quad p_{i,i} = 1-\frac{2D_{2}\tau}{\delta^{2}},\quad p_{i,i+1} = \frac{(2D_{2}+v_{2}\delta)\tau}{2\delta^{2}},\quad i = \mathcal{I}+1,\hdots,N_{s}-1,\\\
p_{i,i-1} = \frac{(2D_{2}-v_{2}\delta)\tau}{\delta^{2}}, \quad p_{i,i} = 1-\frac{(2D_{2}-v_{2}\delta)\tau}{\delta^{2}},\quad i = N_{s},
\end{gather*}
and is a stochastic matrix when $\smash{\tau \leq \min\left\{\frac{\delta^{2}}{2D_{1}},\frac{\delta^{2}}{2D_{2}},\frac{\delta^{2}}{2D_{1}+v_{1}\delta},\frac{\delta^{2}}{2D_{2}-v_{2}\delta},\frac{2\delta^{2}}{2(D_{1}+D_{2})+(v_{2}-v_{1})\delta}\right\}}$ and\\ $\smash{\delta\leq\min\left\{\frac{2D_{1}}{|v_{1}|},\frac{2D_{2}}{|v_{2}|}\right\}}$. Inspecting the transition probabilities, we see that particles move to the left and right with distinct probabilities within each layer that are consistent with those reported for model 4. Separately, the change in diffusivity and advection-velocity at the interface ($i = \mathcal{I}$) produces several possibilities: (i) if $v_{1}$ and $v_{2}$ are both positive or both negative, particles accumulate at the interface if $v_{2}<v_{1}$ and dissipate away from the interface if $v_{2}>v_{1}$ (ii) if $v_{1}$ is positive and $v_{2}$ is negative, particles accumulate at the interface (iii) if $v_{1}$ is negative and $v_{2}$ is positive, particles dissipate away from the interface. All observations are consistent with the continuum model~(model 5 in Table ~\ref{tab:continuum_models}).\bigskip

\subhead{Exact Exponential Discretisation}: Each row of $\mathbf{C}$ sums to zero while all off-diagonal entries of $\mathbf{C}$ are non-negative when $\smash{\delta\leq\min\left\{\frac{2D_{1}}{|v_{1}|},\frac{2D_{2}}{|v_{2}|}\right\}}$. Therefore, the transition matrix $\mathbf{P} = \exp(\tau\mathbf{C})$ is a stochastic matrix when $\smash{\delta\leq\min\left\{\frac{2D_{1}}{|v_{1}|},\frac{2D_{2}}{|v_{2}|}\right\}}$ with no restriction on $\tau$.

\subsection{Model 6: Heterogeneous advection-diffusion with semi-permeable interface}
\label{sec:model6}

\subhead{Discretisation Matrix}: The matrix $\mathbf{C}$ has entries:
\begin{gather*}
c_{i,i} = -\frac{(2D_{1}+v_{1}\delta)}{\delta^{2}}, \quad c_{i,i+1} = \frac{(2D_{1}+v_{1}\delta)}{\delta^{2}},\quad i = 1,\\
c_{i,i-1} = \frac{(2D_{1} - v_{1}\delta)}{2\delta^{2}}, \quad c_{i,i} = -\frac{2D_{1}}{\delta^{2}},\quad c_{i,i+1} = \frac{(2D_{1}+v_{1}\delta)}{2\delta^{2}},\quad i = 2,\hdots,\mathcal{I}-1,\\
c_{i,i-1} = \frac{(2D_{1} - v_{1}\delta)}{\delta^{2}}, \quad c_{i,i} = -\frac{(2D_{1}-v_{1}\delta + 2H\delta)}{\delta^{2}},\quad c_{i,i+1} = \frac{2H}{\delta},\quad i = \mathcal{I},\\
c_{i,i-1} = \frac{2H}{\delta}, \quad c_{i,i} = -\frac{(2D_{2}+v_{2}\delta + 2H\delta)}{\delta^{2}},\quad c_{i,i+1} = \frac{(2D_{2}+v_{2}\delta)}{\delta^{2}},\quad i = \mathcal{I}+1,\\
c_{i,i-1} = \frac{(2D_{2} - v_{2}\delta)}{2\delta^{2}}, \quad c_{i,i} = -\frac{2D_{2}}{\delta^{2}},\quad c_{i,i+1} = \frac{(2D_{2}+v_{2}\delta)}{2\delta^{2}},\quad i = \mathcal{I}+2,\hdots,N_{s}-1,\\\
c_{i,i-1} = \frac{(2D_{2}-v_{2}\delta)}{\delta^{2}}, \quad c_{i,i} = -\frac{(2D_{2}-v_{2}\delta)}{\delta^{2}},\quad i = N_{s}.
\end{gather*}

\subhead{Forward Euler Discretisation}: The transition matrix $\mathbf{P} = \mathbf{I} + \tau\mathbf{C}$ has entries:
\begin{gather*}
p_{i,i} = 1-\frac{(2D_{1}+v_{1}\delta)\tau}{\delta^{2}}, \quad p_{i,i+1} = \frac{(2D_{1}+v_{1}\delta)\tau}{\delta^{2}},\quad i = 1,\\
p_{i,i-1} = \frac{(2D_{1} - v_{1}\delta)\tau}{2\delta^{2}}, \quad p_{i,i} = 1-\frac{2D_{1}\tau}{\delta^{2}},\quad p_{i,i+1} = \frac{(2D_{1}+v_{1}\delta)\tau}{2\delta^{2}},\quad i = 2,\hdots,\mathcal{I}-1,\\
p_{i,i-1} = \frac{(2D_{1} - v_{1}\delta)\tau}{\delta^{2}}, \quad p_{i,i} = 1-\frac{(2D_{1}-v_{1}\delta + 2H\delta)\tau}{\delta^{2}},\quad p_{i,i+1} = \frac{2H\tau}{\delta},\quad i = \mathcal{I},\\
p_{i,i-1} = \frac{2H\tau}{\delta}, \quad p_{i,i} = 1-\frac{(2D_{2}+v_{2}\delta + 2H\delta)\tau}{\delta^{2}},\quad p_{i,i+1} = \frac{(2D_{2}+v_{2}\delta)\tau}{\delta^{2}},\quad i = \mathcal{I}+1,\\
p_{i,i-1} = \frac{(2D_{2} - v_{2}\delta)\tau}{2\delta^{2}}, \quad p_{i,i} = 1-\frac{2D_{2}\tau}{\delta^{2}},\quad p_{i,i+1} = \frac{(2D_{2}+v_{2}\delta)\tau}{2\delta^{2}},\quad i = \mathcal{I}+2,\hdots,N_{s}-1,\\\
p_{i,i-1} = \frac{(2D_{2}-v_{2}\delta)\tau}{\delta^{2}}, \quad p_{i,i} = 1-\frac{(2D_{2}-v_{2}\delta)\tau}{\delta^{2}},\quad i = N_{s},
\end{gather*}
and is a stochastic matrix when $\smash{\tau \leq \min\left\{\frac{\delta^{2}}{2D_{1}},\frac{\delta^{2}}{2D_{2}},\frac{\delta^{2}}{2D_{1}+v_{1}\delta},\frac{\delta^{2}}{2D_{2}-v_{2}\delta},\frac{\delta^{2}}{2D_{1}-v_{1}\delta + 2H\delta},\frac{\delta^{2}}{2D_{2}+v_{2}\delta + 2H\delta}\right\}}$ and $\smash{\delta\leq\min\left\{\frac{2D_{1}}{|v_{1}|},\frac{2D_{2}}{|v_{2}|}\right\}}$. Inspecting the transition probabilities, we see that the only difference between model 5 and model 6 occurs at the interface (sites $\mathcal{I}$ and $\mathcal{I}+1$). As in model 3, the semi-permeable interface produces a barrier between the two layers, where particles transition between layers (from site $\mathcal{I}$ to site $\mathcal{I}+1$ or vice versa) with probability $\frac{2H\tau}{\delta}$. \bigskip

\subhead{Exact Exponential Discretisation}: Each row of $\mathbf{C}$ sums to zero while all off-diagonal entries of $\mathbf{C}$ are non-negative when $\smash{\delta\leq\min\left\{\frac{2D_{1}}{|v_{1}|},\frac{2D_{2}}{|v_{2}|}\right\}}$. Therefore, the transition matrix $\mathbf{P} = \exp(\tau\mathbf{C})$ is a stochastic matrix when $\smash{\delta\leq\min\left\{\frac{2D_{1}}{|v_{1}|},\frac{2D_{2}}{|v_{2}|}\right\}}$ with no restriction on $\tau$.

\subsection{Simulation results}
\begin{figure}[p]
\def\figw{0.45\textwidth}
\centering
\text{Model 1: Homogeneous diffusion}\\[0.5em]
\includegraphics[width=\figw]{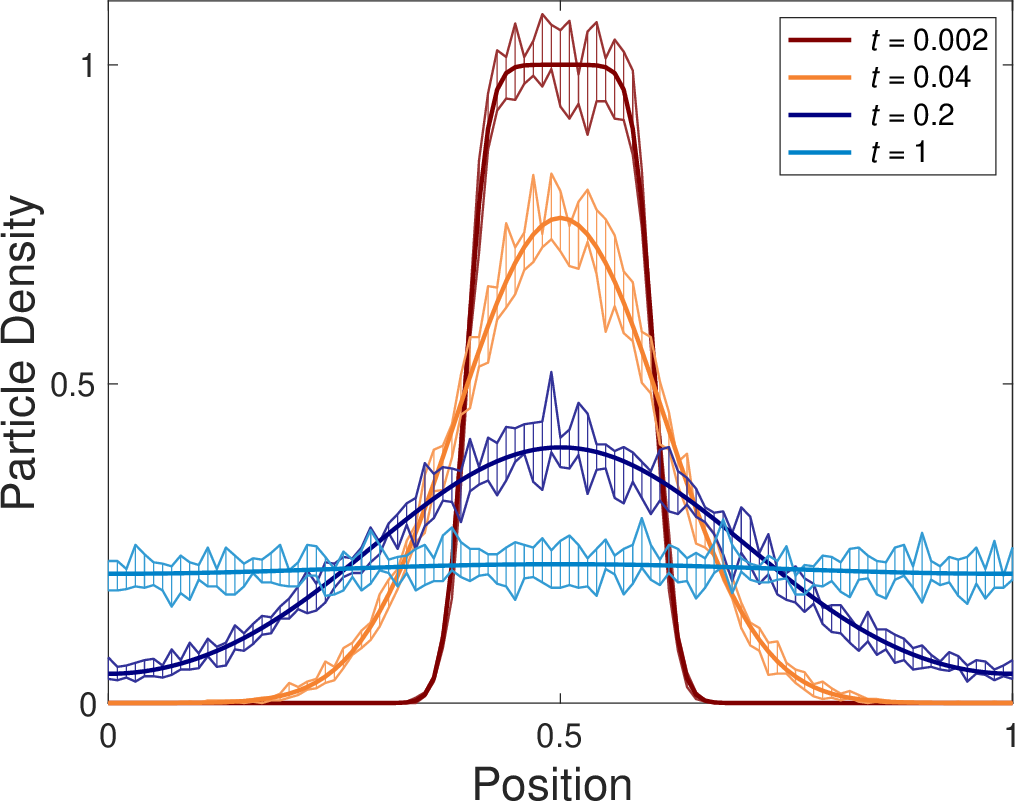}\hfill\includegraphics[width=\figw]{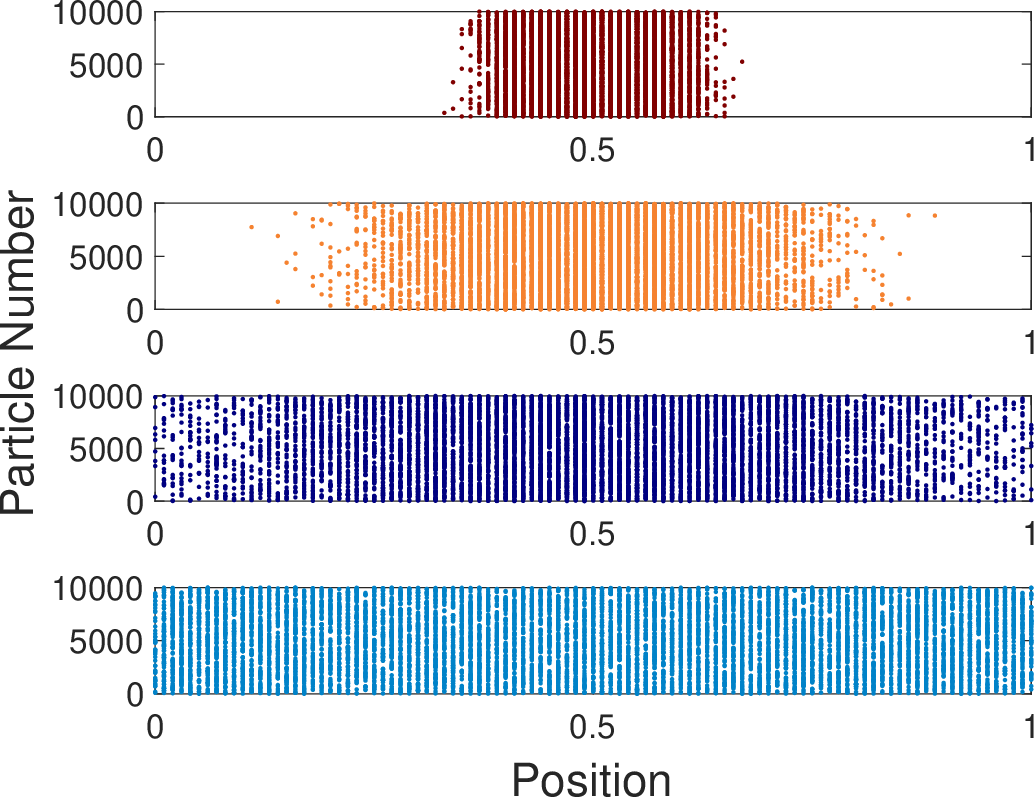}\\[1.5em]
\text{Model 2: Heterogenous diffusion with fully-permeable interface}\\[0.5em]
\includegraphics[width=\figw]{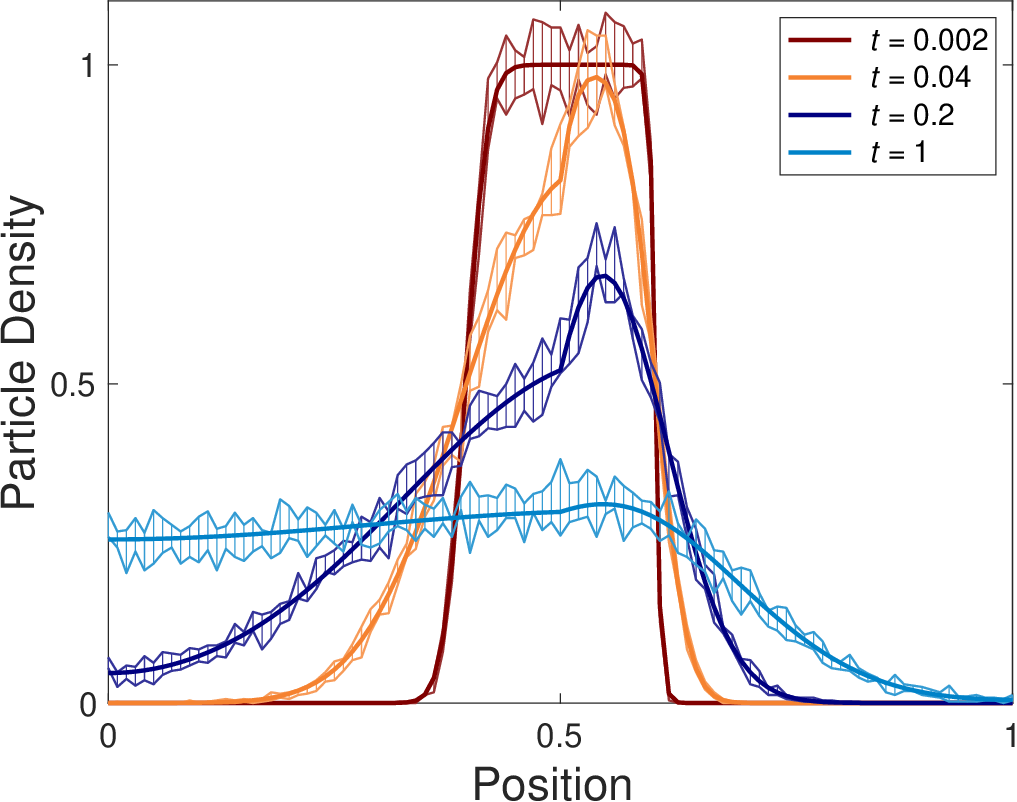}\hfill\includegraphics[width=\figw]{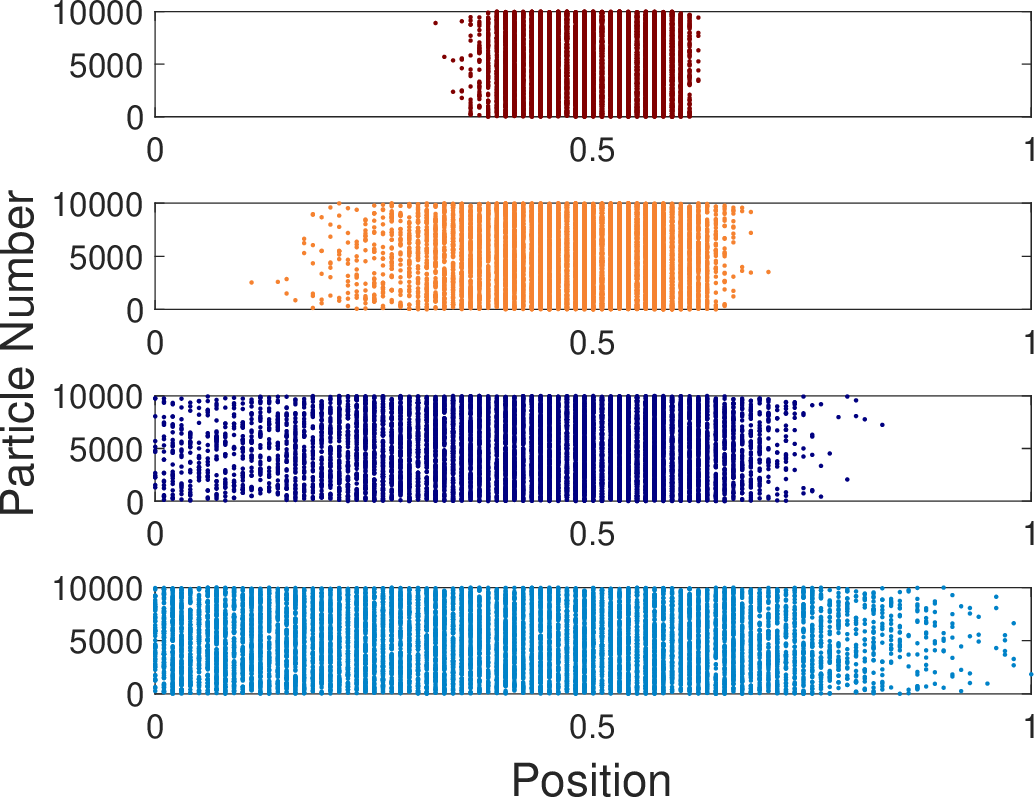}\\[1.5em]
\text{Model 3: Heterogenous diffusion with semi-permeable interface}\\[0.5em]
\includegraphics[width=\figw]{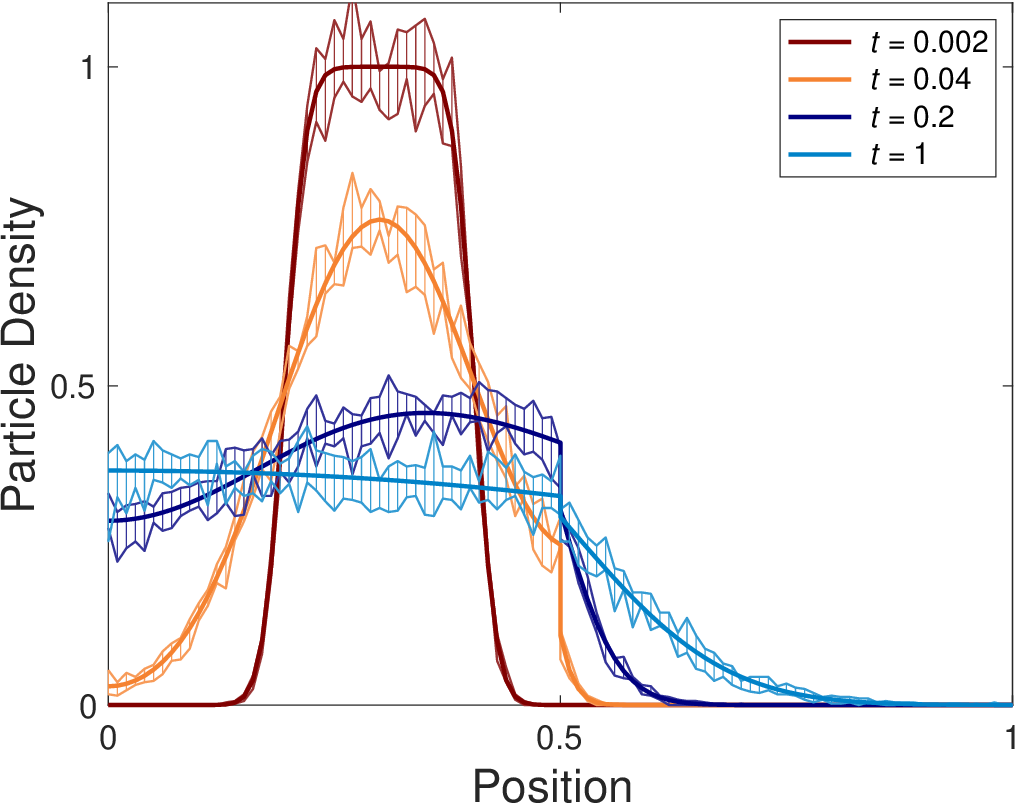}\hfill\includegraphics[width=\figw]{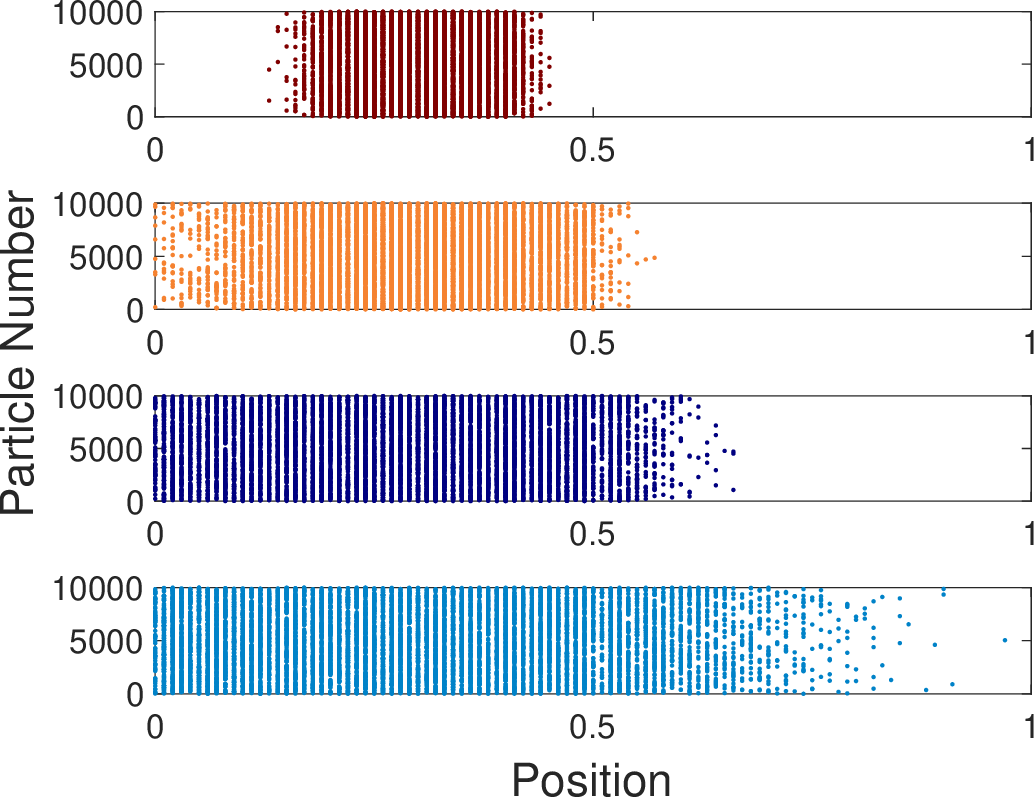}\\[1.5em]
\caption{\textbf{Results for diffusion models (models 1--3)} (left panels) comparison of particle density obtained from the continuum model (continuous line plotting the solution obtained by solving the spatially-discretised continuum model (\ref{eq:ode_U})) and stochastic model (hatched regions enveloping the particle density obtained from $5$ simulations of Algorithm \ref{alg:stochastic_model1} (right panels) results from one simulation of the stochastic model (Algorithm \ref{alg:stochastic_model2}) showing individual particle positions over time. All results correspond to the exact exponential discretisation. The legends indicating time apply across all six panels. Parameters: $f(x) = 1$ if $0.4\leq x\leq 0.6$ and $f(x) = 0$ otherwise (models 1--2), $f(x) = 1$ if $0.2\leq x\leq 0.4$ and $f(x) = 0$ otherwise (model 3), $N_{p} = 10000$, $N_{x} = 101$, $N_{t} = 500$, $T = 1$, $\tau = 0.002$, $L = 1$, $\ell = 0.5$, $\delta = 0.01$, $D = 0.1$ (model 1) $[D_{1},D_{2}] = [0.1,0.01]$ (models 2--3), $H = 0.5$ (model 3).}
\label{fig:diffusion_models}
\end{figure}

\begin{figure}[p]
\def\figw{0.45\textwidth}
\centering
\text{Model 4: Homogeneous advection-diffusion}\\[0.5em]
\includegraphics[width=\figw]{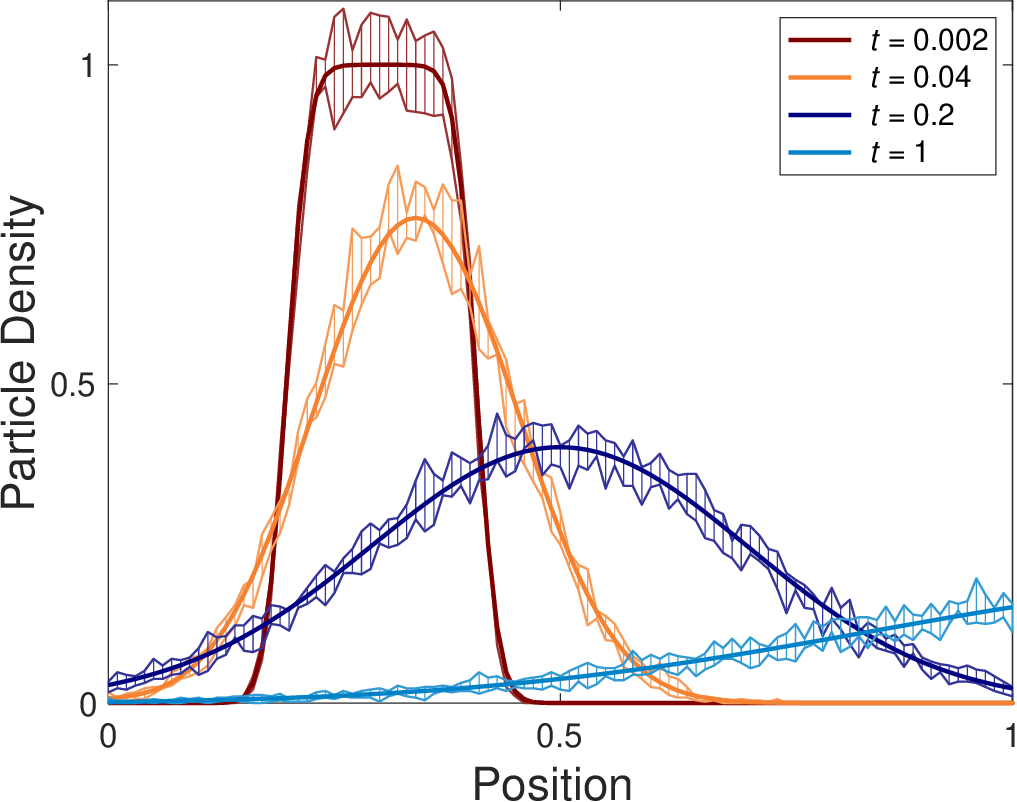}\hfill\includegraphics[width=\figw]{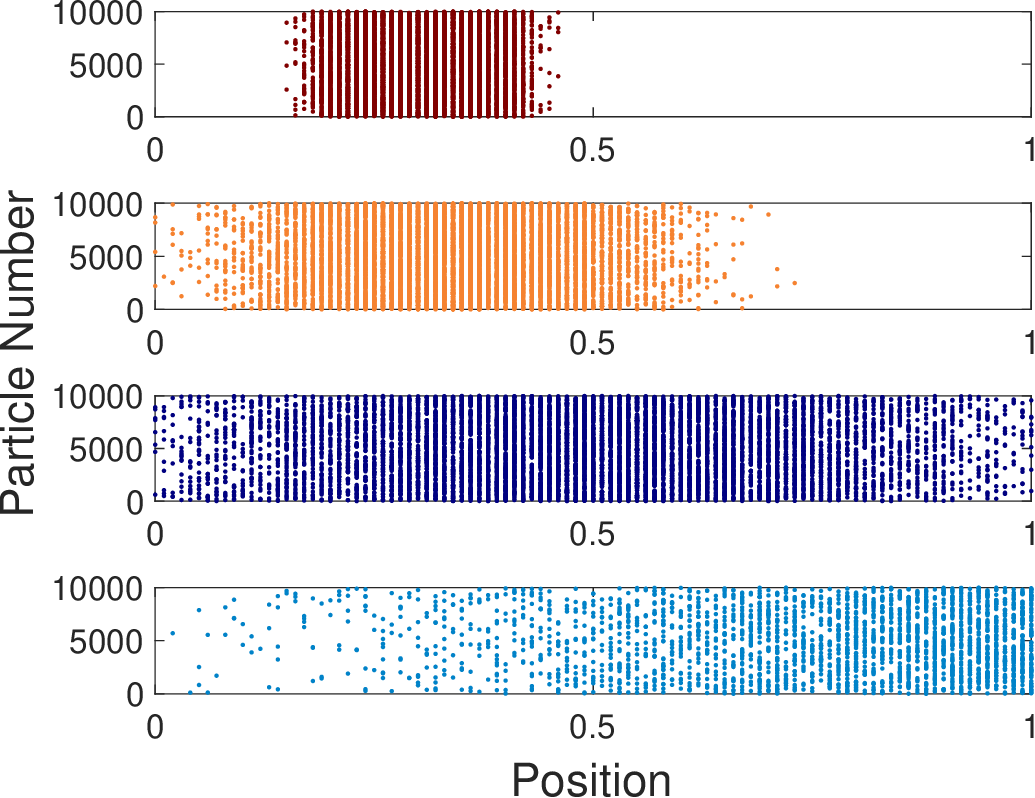}\\[1.5em]
\text{Model 5: Heterogenous advection-diffusion with fully-permeable interface}\\[0.5em]
\includegraphics[width=\figw]{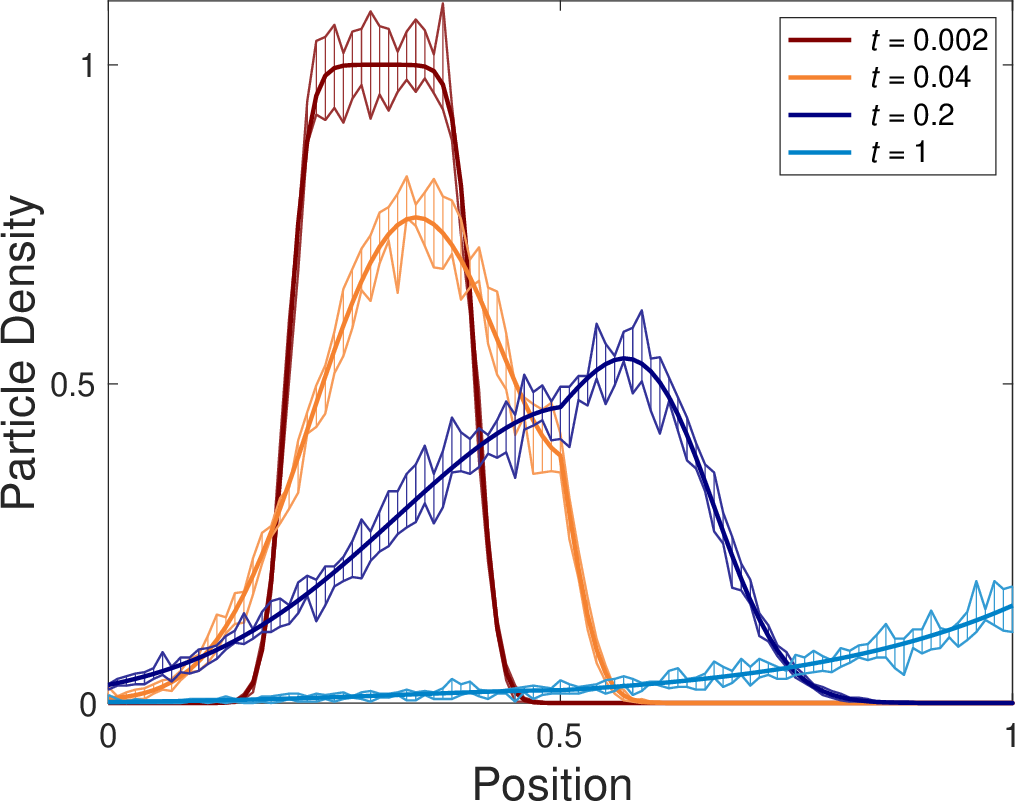}\hfill\includegraphics[width=\figw]{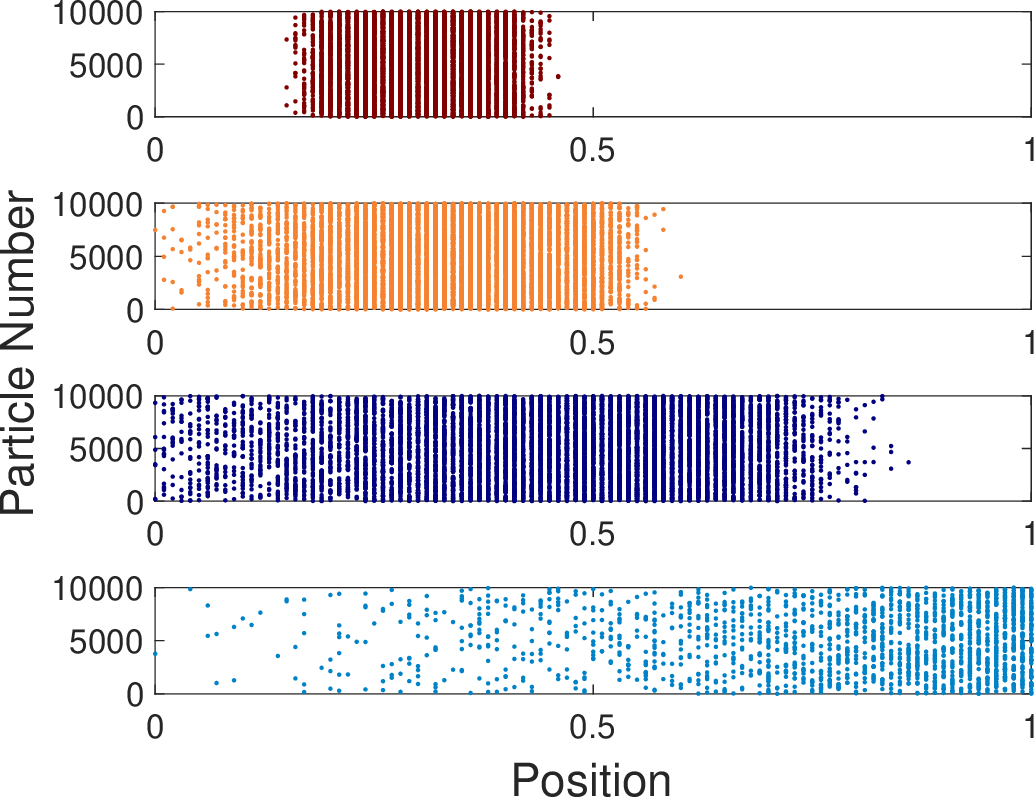}\\[1.5em]
\text{Model 6: Heterogenous advection-diffusion with semi-permeable interface}\\[0.5em]
\includegraphics[width=\figw]{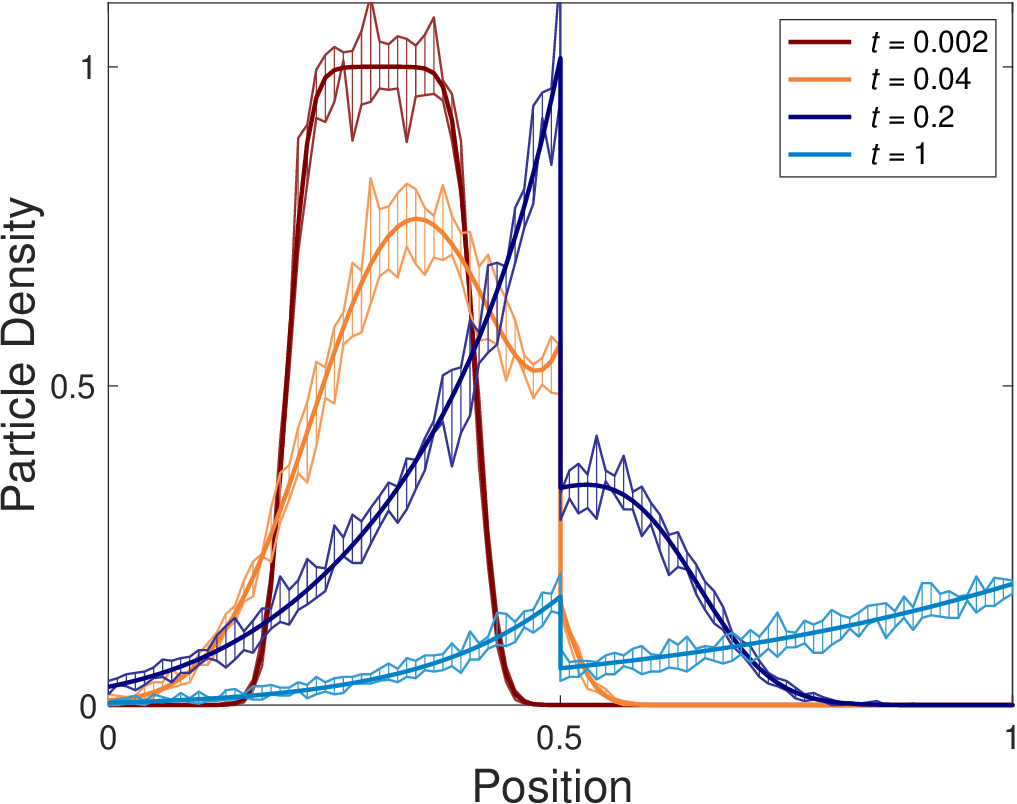}\hfill\includegraphics[width=\figw]{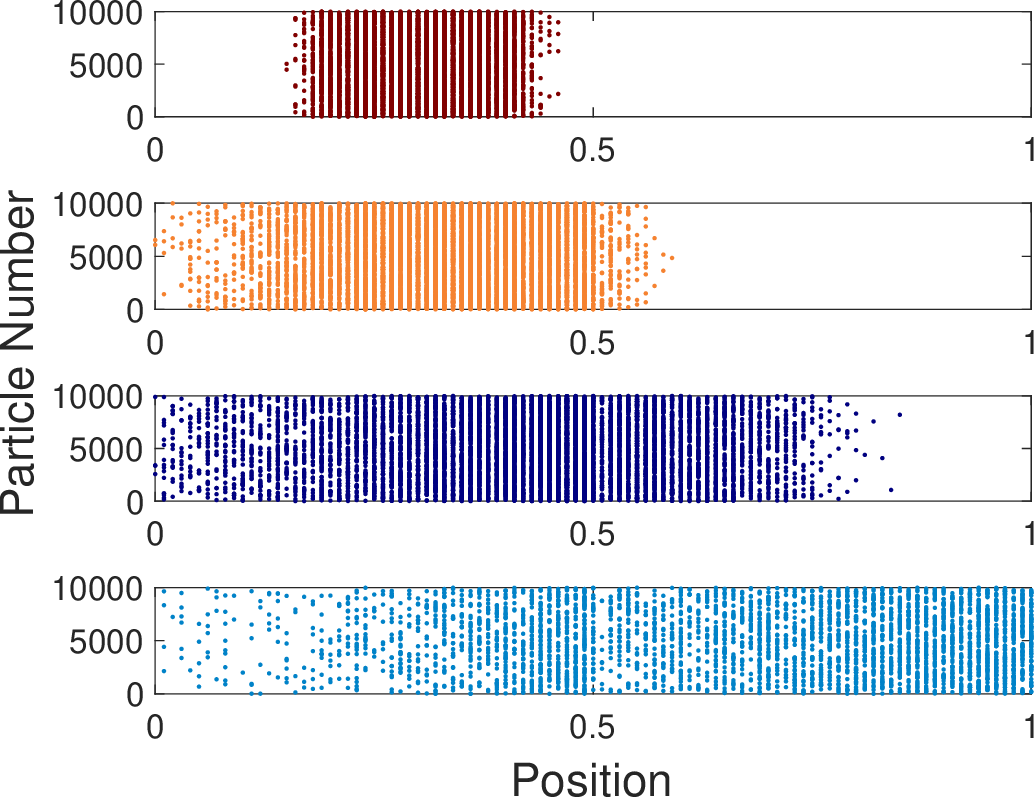}\\[1.5em]
\caption{\textbf{Results for advection-diffusion models (models 4--6)} (left panels) comparison of particle density obtained from the continuum model (continuous line plotting the solution obtained by solving the spatially-discretised continuum model (\ref{eq:ode_U})) and stochastic model (hatched regions enveloping the particle density obtained from $5$ simulations of Algorithm \ref{alg:stochastic_model1} (right panels) results from one simulation of the stochastic model (Algorithm \ref{alg:stochastic_model2}) showing individual particle positions over time. All results correspond to the exact exponential discretisation. The legends indicating time apply across all six panels. Parameters: $f(x) = 1$ if $0.2\leq x\leq 0.4$ and $f(x) = 0$ otherwise, $N_{p} = 10000$, $N_{x} = 501$, $N_{t} = 500$, $T = 1$, $\tau = 0.002$, $L = 5$, $\ell = 0.5$, $\delta = 0.01$, $[D,v] = [0.1,1]$ (model 4) $[D_{1},D_{2},v_{1},v_{2}] = [0.1,0.01,1,1]$ (models 5--6), $H = 0.5$ (model 6).}
\label{fig:advection-diffusion_models}
\end{figure}

Results in Figure \ref{fig:diffusion_models} (diffusion models 1--3) and Figure \ref{fig:advection-diffusion_models} (advection-diffusion models 4--6) provide evidence to support the equivalence of the stochastic and continuum models presented in this work. In these figures, for a specific set of parameters and for all six models, we (i) compare the particle density obtained from the continuum model to the particle density obtained from the stochastic model (ii) provide individual particle positions over time arising from one simulation of the stochastic model. All results correspond to the exact exponential discretisation with $\tau = 0.002$ and $\delta = 0.01$, the latter satisfying the constraint on $\delta$ present for the advection-diffusion models (models 4--6). Similar results were obtained for the forward Euler discretisation (not shown) but this required the time step to be reduced to $\tau = 1/2100$ $(\approx 0.0004762)$ to collectively satisfy the constraints on $\tau$ present for both the diffusion and advection-diffusion models (models 1--6). 
 

From these results, we see that the stochastic model simulations match well with the continuum model solution, with both exhibiting behaviour consistent with diffusion and advection-diffusion in homogeneous and heterogeneous media: 
\begin{itemize}
\item Model 1: particles move to the left and right with equal probability resulting in a symmetric particle density over time. 
\item Model 2: particles move to the left and right with equal probability but this probability is higher in the first layer than the second layer (since $D_{1}>D_{2}$). This results in particles reaching the left boundary at $x=0$ quicker than the right boundary at $x = 1$ and particles accumulating near the interface in the second layer at early times. 
\item Model 3: particles diffusing in the positive $x$ direction reach the semi-permeable interface, where a low transition probability results in a discontinuity in particle density and an accumulation of particles waiting to transition across the interface.
\item Model 4: particles move to the right with higher probability than they move to the left (since $v>0$) resulting in an advection of particles in the positive $x$ direction. 
\item Model 5: particles transition freely between layers and cluster to the right of the interface due to the slower diffusion in the second layer than the first layer. 
\item Model 6: particles are advected in the positive $x$ direction before reaching the semi-permeable interface, where a low transition probability results in a discontinuity in particle density and an accumulation of particles waiting to transition across the interface.
\end{itemize}
Results have been reported for illustrative examples only. MATLAB code available on GitHub (\href{https://github.com/elliotcarr/Carr2024c}{https://github.com/elliotcarr/Carr2024c}) can be used to investigate other parameter choices and view more informative animations of both the continuum and stochastic model simulations.

\section{Conclusions}
\label{sec:conclusions}
In summary, we have derived a suite of equivalent stochastic models (discrete-time discrete-space random walk models) for several standard continuum models of advection-diffusion across a fully- or semi-permeable interface. Our approach involves discretising the continuum model in space and time to derive a Markov chain that governs the transition of particles between spatial lattice sites during each time step. Discretisation in space was carried out using a classical finite volume method while two options were considered for performing the discretisation in time. A forward Euler discretisation led to a tridiagonal transition matrix and hence a stochastic model taking the form of a local (nearest-neighbour) random walk. An exact exponential discretisation led to a dense transition matrix and hence a stochastic model taking the form of a non-local random walk. Simulation results confirmed good agreement between the derived stochastic models and their continuum model counterpart. 

Both time discretisation methods considered in this work have advantages and disadvantages. The forward Euler discretisation provides simple analytical expressions for the transition probabilities (and straightforward analytical insight into the relationship between the stochastic transition probabilities and the continuum model parameters) but requires a constraint on the time step size to ensure non-negative transition probabilities. The exact exponential discretisation provides transition probabilities defined numerically via a matrix exponential (no analytical insight into the relationship between the stochastic transition probabilities and the continuum model parameters) but does not require a constraint on the time step size to ensure non-negative transition probabilities. Comparably, both methods require the same constraints on the spatial step size for advection-diffusion problems (models 4--6 in Table \ref{tab:continuum_models}).

Finally, it is important to note that the stochastic models developed in this work are limited to the specific continuum models considered. While multiple layers can be accommodated by a straightforward amalgamation of the results presented, extending the analysis to lattice-free stochastic models, two or three dimensional continuum models or non-uniform spatial steps would yield different transition probabilities but could be interesting to pursue in the future.


\appendix 
\section*{Appendix}
\renewcommand{\thesubsection}{A.\arabic{subsection}}
\renewcommand{\theequation}{A.\arabic{equation}}
\renewcommand{\thealgorithm}{A.\arabic{algorithm}}

\subsection{Alternative stochastic model algorithm}
\label{app:stochastic_model2}
\medskip\noindent
\algbox{
\begin{algorithm}[Stochastic Model]\mbox{}\\
\label{alg:stochastic_model2}
\emph{
\begin{tabular}{l}
$U_{i,0} = f(x_{i})$  for all $i = 1,\hdots,N_{s}$ \comment{particle density at lattice site $i$ and time $t=0$}\\
$S_{p} = \sum_{i=1}^{N_{s}}U_{i,0}V_{i}/N_{p}$ \comment{scaling constant}\\
$N_{i,0} = \text{round}(U_{i,0}V_{i}/S_{p})$ for all $i = 1,\hdots,N_{s}$ \comment{initial number of particles at site $i$}\\
$N_{p} = \sum_{i=1}^{N_{s}} N_{i,0}$ \comment{corrected number of particles}\\
$x_{k,0} = i$ for all $k = \sum_{m=1}^{i-1}N_{m,0}+1,\hdots,\sum_{m=1}^{i}N_{m,0}$ and $i = 1,\hdots,N_{s}$ \comment{initial site of particle $k$}\\ 
$P_{i,0} = 0$ and $P_{i,j} = \sum_{m=1}^{j} p_{i,m}$ for all $i=1,\hdots,N_{s}$ and $j=1,\hdots,N_{s}$ \comment{cumulative probabilities}\\
for $n = 1,2,\hdots,N_{t}$ \comment{loop over time steps}\\
\qquad for $k = 1,2,\hdots,N_{p}$ \comment{loop over number of particles}\\
\qquad \qquad Sample $r \sim \text{Uniform}(0,1)$ \comment{uniform random number in $[0,1]$}\\
\qquad \qquad $i = x_{k,n-1}$ \comment{lattice site for particle $k$ at $t = t_{n-1}$}\\
\qquad \qquad Find $j$ such that $r\in(P_{i,j-1},P_{i,j})$ \comment{particle moves from site $i$ to site $j$ at $t = t_{n}$}\\
\qquad \qquad $x_{k,n} = j$ \comment{update lattice site for particle $k$ at $t = t_{n}$}\\
\qquad end\\
\qquad $N_{i,n} = \sum_{k=1}^{N_{p}} [x_{k,n} = i]$ for all $i = 1,\hdots,N_{s}$ \comment{number of particles at site $i$ and time $t = t_{n}$}\\
\qquad $U_{i,n} = N_{i,n}S_{p}/V_{i}$ for all $i = 1,\hdots,N_{s}$ \comment{particle density at lattice site $i$ and time $t = t_{n}$}\\
end\\
\end{tabular}
}
\end{algorithm}}
Note: the expression $[x_{k,n} = i]$ returns $1$ if $x_{k,n} = i$ and $0$ if $x_{k,n}\neq i$.

\subsection{Finite volume discretisations}
\label{app:fvm_discretisations}
In this appendix, we provide the form of the discretisation matrix $\mathbf{A}$ appearing in the system of differential equations (\ref{eq:ode_U}) for each of the advection-diffusion models (models 4--6) considered in this work (see Table \ref{tab:continuum_models}). For the diffusion models (models 1--3), $\mathbf{A} = \mathbf{C}$, where $\mathbf{C}$ is defined in sections \ref{sec:model1}--\ref{sec:model3}, respectively. In the definitions below, $a_{i,j}$ denotes the entry of $\mathbf{A}$ in row $i$ and column $j$.\bigskip

\noindent Model 4: Homogeneous advection-diffusion
\begin{gather*}
a_{i,i} = -\frac{(2D+v\delta)}{\delta^{2}}, \quad a_{i,i+1} = \frac{(2D-v\delta)}{\delta^{2}},\quad i = 1,\\
a_{i,i-1} = \frac{(2D+v\delta)}{2\delta^{2}}, \quad a_{i,i} = -\frac{2D}{\delta^{2}},\quad a_{i,i+1} = \frac{(2D-v\delta)}{2\delta^{2}},\quad i = 2,\hdots,N_{s}-1,\\
a_{i,i-1} = \frac{(2D+v\delta)}{\delta^{2}}, \quad a_{i,i} = -\frac{(2D-v\delta)}{\delta^{2}},\quad i = N_{s}.
\end{gather*}

\noindent Model 5: Heterogeneous advection-diffusion with fully-permeable interface
\begin{gather*}
a_{i,i} = -\frac{(2D_{1}+v_{1}\delta)}{\delta^{2}}, \quad a_{i,i+1} = \frac{(2D_{1}-v_{1}\delta)}{\delta^{2}},\quad i = 1,\\
a_{i,i-1} = \frac{(2D_{1} + v_{1}\delta)}{2\delta^{2}}, \quad a_{i,i} = -\frac{2D_{1}}{\delta^{2}},\quad a_{i,i+1} = \frac{(2D_{1}-v_{1}\delta)}{2\delta^{2}},\quad i = 2,\hdots,\mathcal{I}-1,\\
a_{i,i-1} = \frac{(2D_{1} + v_{1}\delta)}{2\delta^{2}}, \quad a_{i,i} = -\frac{[2(D_{1}+D_{2}) + (v_{2}-v_{1})\delta]}{2\delta^{2}},\quad a_{i,i+1} = \frac{(2D_{2}-v_{2}\delta)}{2\delta^{2}},\quad i = \mathcal{I},\\
a_{i,i-1} = \frac{(2D_{2} + v_{2}\delta)}{2\delta^{2}}, \quad a_{i,i} = -\frac{2D_{2}}{\delta^{2}},\quad a_{i,i+1} = \frac{(2D_{2}-v_{2}\delta)}{2\delta^{2}},\quad i = \mathcal{I}+1,\hdots,N_{s}-1,\\\
a_{i,i-1} = \frac{(2D_{2}+v_{2}\delta)}{\delta^{2}}, \quad a_{i,i} = -\frac{(2D_{2}-v_{2}\delta)}{\delta^{2}},\quad i = N_{s}.
\end{gather*}

\noindent Model 6: Heterogeneous advection-diffusion with semi-permeable interface
\begin{gather*}
a_{i,i} = -\frac{(2D_{1}+v_{1}\delta)}{\delta^{2}}, \quad a_{i,i+1} = \frac{(2D_{1}-v_{1}\delta)}{\delta^{2}},\quad i = 1,\\
a_{i,i-1} = \frac{(2D_{1} + v_{1}\delta)}{2\delta^{2}}, \quad a_{i,i} = -\frac{2D_{1}}{\delta^{2}},\quad a_{i,i+1} = \frac{(2D_{1}-v_{1}\delta)}{2\delta^{2}},\quad i = 2,\hdots,\mathcal{I}-1,\\
a_{i,i-1} = \frac{(2D_{1} + v_{1}\delta)}{\delta^{2}}, \quad a_{i,i} = -\frac{(2D_{1}-v_{1}\delta + 2H\delta)}{\delta^{2}},\quad a_{i,i+1} = \frac{2H}{\delta},\quad i = \mathcal{I},\\
a_{i,i-1} = \frac{2H}{\delta}, \quad a_{i,i} = -\frac{(2D_{2}+v_{2}\delta + 2H\delta)}{\delta^{2}},\quad a_{i,i+1} = \frac{(2D_{2}-v_{2}\delta)}{\delta^{2}},\quad i = \mathcal{I}+1,\\
a_{i,i-1} = \frac{(2D_{2} + v_{2}\delta)}{2\delta^{2}}, \quad a_{i,i} = -\frac{2D_{2}}{\delta^{2}},\quad a_{i,i+1} = \frac{(2D_{2}-v_{2}\delta)}{2\delta^{2}},\quad i = \mathcal{I}+2,\hdots,N_{s}-1,\\\
a_{i,i-1} = \frac{(2D_{2}+v_{2}\delta)}{\delta^{2}}, \quad a_{i,i} = -\frac{(2D_{2}-v_{2}\delta)}{\delta^{2}},\quad i = N_{s}.
\end{gather*}

\footnotesize
\setlength{\bibsep}{1pt plus 0.3ex}

\end{document}